\documentclass[twoside, onecolumn,11pt]{article}
\usepackage{extsizes}
\usepackage[super,sort&compress,comma]{natbib} 
\usepackage[version=4]{mhchem}
\usepackage[left=1.5cm, right=1.5cm, top=1.785cm, bottom=2.0cm]{geometry}
\usepackage{balance}
\usepackage{times,mathptmx}
\usepackage{sectsty}
\usepackage{graphicx} 
\usepackage{lastpage}
\usepackage[format=plain,justification=justified,singlelinecheck=true,font={stretch=1.125,small,sf},labelfont=bf,labelsep=space]{caption}
\usepackage{float}
\usepackage{fancyhdr}
\usepackage{fnpos}
\usepackage[english]{babel}
\addto{\captionsenglish}{%
  
}
\usepackage{array}
\usepackage{droidsans}
\usepackage{charter}
\usepackage[T1]{fontenc}
\usepackage[usenames,dvipsnames]{xcolor}
\usepackage{setspace}
\usepackage[compact]{titlesec}
\usepackage{hyperref}

\usepackage{epstopdf}

\definecolor{cream}{RGB}{222,217,201}
\newcommand{\red}[1]{{\color{black} #1}}
\newcommand{\blue}[1]{{\color{black} #1}}

\newcommand{\magenta}[1]{{\color{black} #1}}

\usepackage{stackengine}

\usepackage{titlesec}
\titlelabel{\thetitle.\quad}

\usepackage{etoolbox}

\makeatletter
\let\oldcite\cite
\pretocmd{\listoffigures}{\def\cite{\ignorespaces\@gobble}}{}{}
\apptocmd{\listoffigures}{\let\cite\oldcite}{}{}
\pretocmd{\listoftables}{\def\cite{\ignorespaces\@gobble}}{}{}
\apptocmd{\listoftables}{\let\cite\oldcite}{}{}
\makeatother

\begin{document}


\makeFNbottom
\makeatletter
\renewcommand\LARGE{\@setfontsize\LARGE{15pt}{17}}
\renewcommand\Large{\@setfontsize\Large{12pt}{14}}
\renewcommand\large{\@setfontsize\large{10pt}{12}}
\renewcommand\footnotesize{\@setfontsize\footnotesize{7pt}{10}}
\makeatother

\renewcommand{\thefootnote}{\fnsymbol{footnote}}
\renewcommand\footnoterule{\vspace*{1pt}%
\color{cream}\hrule width 3.5in height 0.4pt \color{black}\vspace*{5pt}} 
\setcounter{secnumdepth}{5}

\makeatletter 
\renewcommand\@biblabel[1]{#1}            
\renewcommand\@makefntext[1]%
{\noindent\makebox[0pt][r]{\@thefnmark\,}#1}
\makeatother 
\renewcommand{\figurename}{\small{Figure}}
\sectionfont{\sffamily\Large}
\subsectionfont{\normalsize}
\subsubsectionfont{\bf}
\setstretch{1.125} 
\setlength{\skip\footins}{0.8cm}
\setlength{\footnotesep}{0.25cm}
\setlength{\jot}{10pt}
\titlespacing*{\section}{0pt}{4pt}{4pt}
\titlespacing*{\subsection}{0pt}{15pt}{1pt}

\makeatletter 
\newlength{\figrulesep} 
\setlength{\figrulesep}{0.5\textfloatsep} 

\newcommand{\topfigrule}{\vspace*{-1pt}%
\noindent{\color{cream}\rule[-\figrulesep]{\columnwidth}{1.5pt}} }

\newcommand{\botfigrule}{\vspace*{-2pt}%
\noindent{\color{cream}\rule[\figrulesep]{\columnwidth}{1.5pt}} }

\newcommand{\dblfigrule}{\vspace*{-1pt}%
\noindent{\color{cream}\rule[-\figrulesep]{\columnwidth}{1.5pt}} }

\makeatother

\begin{center}
\noindent\LARGE\textbf{Exploring $\rm Mg^{2+}$ and $\rm Ca^{2+}$ Conductors Via Solid-State Metathesis Reactions
}\\
\noindent\LARGE\centering{}
\end{center}

\noindent\large{Titus Masese,\textit{$^{a, c}$} Godwill Mbiti Kanyolo,\textit{$^{a, b}$} Yoshinobu Miyazaki,\textit{$^d$} Shintaro Tachibana,\textit{$^e$} Keigo Kubota,\textit{$^{a}$} Naoya Ishida,\textit{$^{a}$} Kohei Tada,\textit{$^{a, f}$} Hiroyuki Ozaki,\textit{$^{a}$} Toyoki Okumura,\textit{$^{a}$} Sahori Takeda,\textit{$^{a}$} Sachio Komori,\textit{$^g$} Tomoyasu Taniyama,\textit{$^g$} Yuki Orikasa,\textit{$^e$} and Tomohiro Saito\textit{$^d$} }\\

\noindent{\textit{$^{a}$Research Institute of Electrochemical Energy, National Institute of Advanced Industrial Science and Technology (AIST), 1-8-31 Midorigaoka, Ikeda, Osaka 563-8577, Japan.} Email: titus.masese@aist.go.jp; gm.kanyolo@aist.go.jp\\%
\textit{$^{b}$Department of Engineering Science, The University of Electro-Communications, 1-5-1 Chofugaoka, Chofu, Tokyo 182-8585, Japan. } \\
\textit{$^{c}$AIST-Kyoto University Chemical Energy Materials Open Innovation Laboratory (ChEM-OIL), Sakyo-ku, Kyoto 606-8501, Japan.}\\
\textit{$^{d}$Tsukuba Laboratory, Sumika Chemical Analysis Service (SCAS), Ltd.,Tsukuba, Ibaraki 300–3266, Japan.}\\
\textit{$^{e}$Graduate School of Life Sciences, Ritsumeikan University, 1–1–1 Noji-higashi, Kusatsu, Shiga 525–8577, Japan.}\\
\textit{$^{f}$Department of Materials Engineering Science, Graduate School of Engineering Science, Osaka University, 1-3 Machikaneyama-cho, Toyonaka, Osaka 560–8531, Japan.}\\
\textit{$^{g}$Department of Physics, Nagoya University, Furo-cho, Chikusa-ku, Nagoya 464–8602, Japan.}\\

\noindent\normalsize{
\textbf{Magnesium and calcium batteries offer promising energy storage solutions characterised by cost-effective\\ness, safety, and high energy density. However, the scarcity of viable electrode and electrolyte materials vastly hinders their advancement. This study utilises solid-state metathetical reactions involving predominantly chalcogen- and pnictogen-based honeycomb layered oxides with alkaline-earth halides/nitrates to synthesise $\rm Mg^{2+}$- and $\rm Ca^{2+}$-based materials previously achievable only under high-temperature/high-pressure conditions, as well as new metastable materials with unique crystal versatility. Particularly, we employ metathetical reactions involving $\rm Li_4MgTeO_6$, $\rm Na_2Mg_2TeO_6$, and $\rm Na_4MgTeO_6$ with \magenta {$\rm MgCl_2$\,/ $\rm MgSO_4$\,/\\\,$\rm Mg(NO_3)_2$·$\rm 6H_2O$ or $\rm Ca(NO_3)_2$·$\rm 4H_2O$ / $\rm CaCl_2$·$\rm 2H_2O$} at temperatures not exceeding 500 $^\circ$C to produce $\rm Mg_3TeO_6$ polymorphs, ilmenite-type $\rm CaMg_2TeO_6$\,/\,$\rm Mg_2CaTeO_6$, and double perovskite-type $\rm Ca_2MgTeO_6$. Thus, we demonstrate that these materials, conventionally requiring gigascale pressures or high temperatures (>1000$^\circ$C) for their proper synthesis, are now readily accessible at ambient pressure and considerably lower temperatures. Meanwhile, despite sub-optimal pellet densities, the synthesised ilmenite-type \magenta {$\rm Mg_3TeO_6$ (high-pressure polymorph)} and double perovskite-type ${\rm Ca}_2M{\rm TeO_6}$ ($M = \rm Mg, Ca, Zn$) materials exhibit remarkable bulk ionic conductivity at room temperature, marking them as promising compositional spaces for exploring novel $\rm Mg^{2+}$ and $\rm Ca^{2+}$ conductors. Furthermore, this study extends the applicability of metathetical reactions to attain Mg- or Ca-based  antimonates, ruthenates, titanates, phosphates, and silicates, thus opening avenues to novel high-entropy multifunctional nanomaterial platforms with utility in energy storage and beyond.}}

\begin{figure*}[!t]
 \centering
 \includegraphics[width=0.83\columnwidth]{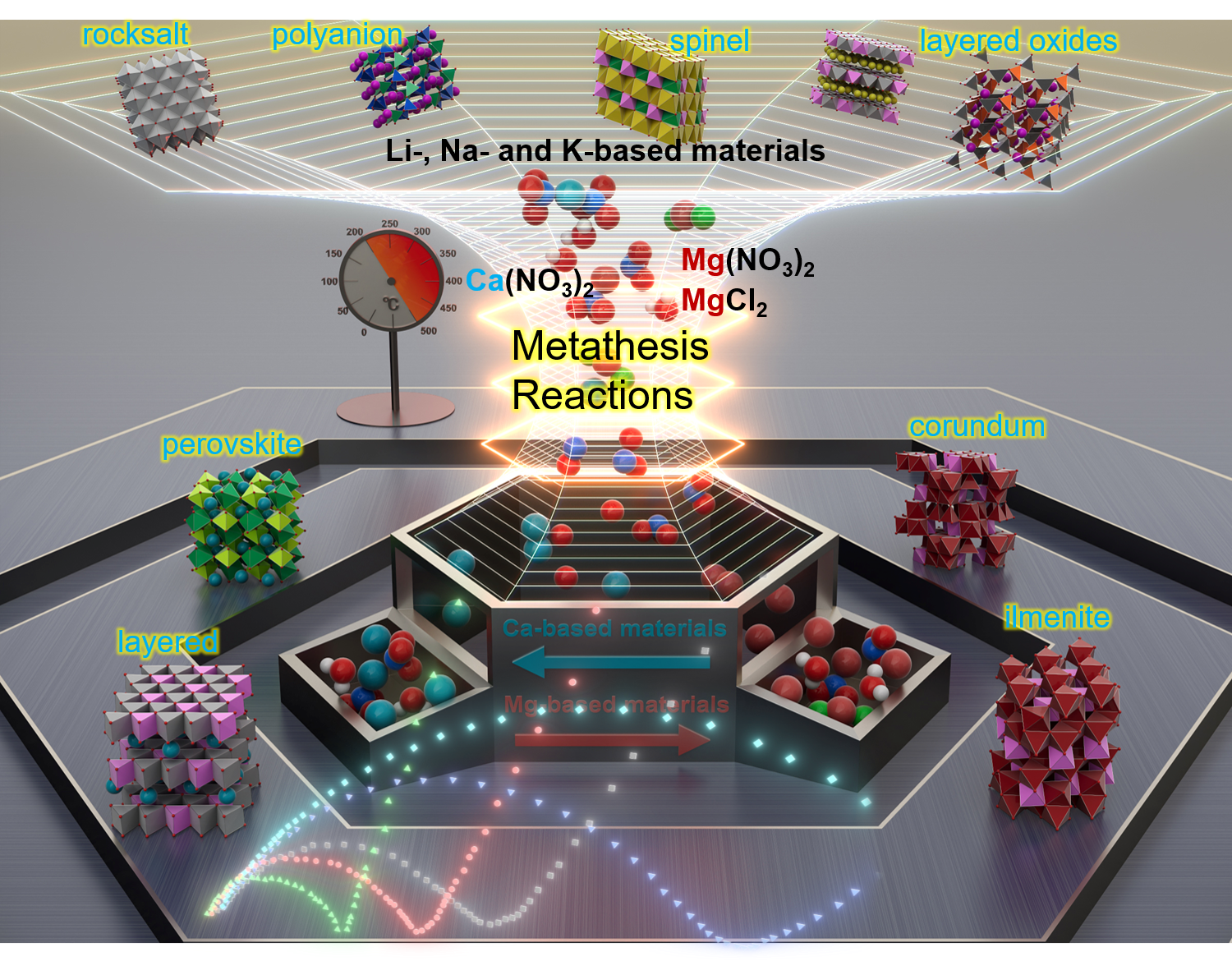}
\end{figure*}


\renewcommand*\rmdefault{bch}\normalfont\upshape
\rmfamily
\section*{}
\vspace{-1cm}


\newpage


\newpage


\newpage 


\newpage

\section{\label{Section: Introduction} Introduction}
The dawn of nanotechnology has ushered in a new paradigm of nanomaterials, characterised by a rich tapestry of exceptional chemical properties and physical functionalities. Specifically, honeycomb layered oxides based on pnictogens \blue {(group 15 elements)} and chalcogens \blue {(group 16 elements)} have surfaced, showcasing intricate crystal chemistry alongside unique electromagnetic and topological phenomena.\cite {kanyolo2023honeycomb, kanyolo2023advances, kanyolo2021honeycomb, kanyolo2023pseudo} These oxide materials hold significant value and elicit diverse interest across multiple disciplines, ranging from condensed matter to materials science, with promising applications in engineering, electronics, and energy storage.\cite {kanyolo2023honeycomb, kanyolo2023advances} This category of layered oxides predominantly encompasses mobile alkali metal cations (such as $\rm K, Na, Li$) or coinage metal cations (such as $\rm Ag$ or $\rm Cu$), interposed between layers of primarily transition metals ($\rm Zn, Cu, Ni, Co, Fe, Mn$, or $\rm Cr$) arranged in a honeycomb framework encircling chalcogens (such as $\rm Te$) or pnictogens ($\rm Bi, Sb $, or $\rm As$) coordinated with oxygen atoms, thereby presenting an intriguing compositional space. 

\blue {The syntheses of calcium- and magnesium-based analogue compositions, for instance, through the conventional solid-state ceramics route, often necessitates high temperatures or pressures to obtain the desired target materials.\cite {selb2019crystal, burckhardt1982} In certain cases, this approach solely yields impurities or non-equilibrium intermediates.} \red{Thus, t}o extend the compositional breadth of \red{the aforementioned class of materials} \red{to encompass} those that include mobile alkaline-earth metals \red{for} multivalent energy storage systems such as $\rm Ca$ and $\rm Mg$ batteries,\cite {arroyo2019achievements, saha2014rechargeable} judicious synthetic routes become imperative. 
\red{Particularly, s}olid-state metathesis (double-ion-exchange) reactions are emerging as highly effective synthetic pathways for diverse technologically significant inorganic materials, including pnictides\cite {treece1994metathetical}, silicides\cite {parkin1996solid}, carbides\cite {nartowski1998rapid}, borides\cite {parkin1996solid}, metal halides \cite {ugemuge2011preparation}, nitrides\cite {gillan1994rapid}, phosphides,\cite {jarvis2000self} chalcogenides\cite {miura2020selective, parkin1996solid, bonneau1992solid, martinolich2016} and oxides\cite {mandal2004rocksalt, mandal2005new, parkin1996solid, gopalakrishnan2000transformations, haraguchi2018magnetic, haraguchi2019frustrated, sivakumar2004}. These reactions are predominantly instigated by the formation of stoichiometric equivalents of a thermodynamically stable byproduct \red{such as a halide or a nitrate} with high lattice energy,\cite {parkin1996solid} in conjunction with the targeted material. 
\red{For instance, the} solid-state metathetic reaction of the alkaline-earth halides, \blue {sulphates} or nitrates with honeycomb-layered oxide precursor compositions, $A_aM_mL_l\rm O_6$\cite {kanyolo2023honeycomb, kanyolo2023advances} where $A = \rm Li, Na, K, Cu, Ag, {\it etc}.$, $M$ = $s$-block metal atoms \textit {e.g.} $\rm Ca, Mg$ or transition metal atoms (\textit {e.g.}, $ \rm Ni, Co, Cu, Zn,$ \textit {etc}.), $L$ = chalcogen, pnictogen or transition metal atoms (such as $\rm Ru$, \blue {or $\rm W, Ta, Nb,$ \textit {etc}.(for disordered (rocksalt) derivatives)}) and $0 < a \leq4$, $0 < m \leq2$, $0 < l \leq 1$, \red{can take the form}:
\begin{subequations}\label{exemplified_eqs}
\begin{align}
    \magenta{\,{\rm {\it A_aM_mL_l}O_6} + \,{\rm \red{\frac{\mathit a}{2}}\,(MgCl_2\,\,{\rm or}\,\,MgSO_4)} \ce{->[\stackon{}{250--500 $^\circ$C}][\stackunder{}{ambient or vacuum pressures}]}
    {\rm ({\it a\,A}Cl\,\,{\rm or}\,\,{\rm {\frac{\mathit a}{2}}{\it A}_2SO_4})} + \,{\rm Mg{\it _{a/2}M_mL_l}O_6}\red{,}}
\\ \,{\rm {\it A_aM_mL_l}O_6} + \,{\rm \magenta{\frac{\mathit a}{2}}\,Mg(NO_3)_2 \cdot \rm 6H_2O} \magenta{\ce{->[\stackon{}{250 -- 500 $^\circ$C}][\stackunder{}{ambient pressures}]}} {\rm {\it a\,A}NO_3} + \,{\rm Mg{\it _{a/2}M_mL_l}O_6} + \,{\rm \magenta{\frac{\mathit a}{2}}\,6H_2O},
\\ \,{\rm {\it A_aM_mL_l}O_6} + \,{\rm \magenta{\frac{\mathit a}{2}}\,Ca(NO_3)_2 \cdot \rm 4H_2O} \magenta{\ce{->[\stackon{}{250 -- 500 $^\circ$C}][\stackunder{}{ambient pressures}]} {\rm {\it a\,A}NO_3} + \,{\rm Ca{\it _{a/2}M_mL_l}O_6} + \,{\rm \red{\frac{\mathit a}{2}}\,4H_2O}},
\\ \,{\rm {\it A_aM_mL_l}O_6} + \,{\rm \magenta{\frac{\mathit a}{2}}\,CaCl_2 \cdot \rm 2H_2O} \magenta{\ce{->[\stackon{}{250 -- 500 $^\circ$C}][\stackunder{}{ambient pressures}]} {\rm {\it a\,A}Cl} + \,{\rm Ca{\it _{a/2}M_mL_l}O_6} + \,{\rm \red{\frac{\mathit a}{2}}\,2H_2O}}.
\end{align}
\end{subequations}

\blue {The substantial lattice energy of the halide or nitrate coproduct salt ($\rm {\it A}Cl$ or $\rm {\it A}NO_3$) can rationalise its pivotal role in propelling the reaction forward,}\cite {parkin1996solid, bonneau1992solid, nartowski1998rapid} thereby facilitating the formation of the desired product. Whilst select metathesis reactions may exhibit spontaneous initiation,\cite {parkin1996solid} the majority necessitate additional energy (accomplished via an external heat source or application of microwave electromagnetic radiation) to surmount a sufficient reaction energy barrier, crucial for propagating a synthesis wave.\blue {\cite {parkin1996solid, bonneau1992solid, nartowski1998rapid}}
Solid-state metathesis reactions then proceed expeditiously, harnessing the liberated enthalpy to elevate the temperature of the precursors, typically surpassing the boiling point of the alkali metal or alkaline-earth metal halide/nitrate. The resulting products, obtained upon completion of the reaction, are readily isolated by trituration with appropriate solvents (such as distilled water, ethanol, methanol, \textit {etc}.), effectively eliminating the halide/nitrate salts and other metathesis byproducts. Consequently, synthesis through solid-state metathesis reactions offers a straightforward and highly scalable low-temperature methodology to access both new and pre-existing functional materials, otherwise achievable solely at high temperatures.\cite {parkin1996solid} Moreover, the resultant products exhibit distinctive and meticulously controlled nanomorphology.\cite {mandal2005new, parhi2008synthesis, parhi2009novel, mandal2004rocksalt, kramer2009microwave, miura2020selective, parhi2008novel}

\red{In this study}, we \red{demonstrate that} solid-state metathesis reactions involving the broad class of chalcogen- and pnictogen-based honeycomb layered oxides and appropriate alkaline-earth metal chloride/nitrates (as exemplified in \red{equation (\ref{exemplified_eqs})}) \red{is} a convenient route to synthesise not only pre-existing materials (attainable only at \textit{high temperatures} or {\it pressures}), but also new $\rm Ca $- and $\rm Mg $-based functional materials at \red{conventionally lower} temperatures and ambient pressure (\textbf {Supplementary Figure 1}). Solid-state metathesis reaction of, for instance, layered $\rm Na_4MgTeO_6$ and $\rm Na_2Mg_2TeO_6$ with \magenta {$\rm Ca(NO_3)_2 \cdot \rm 4H_2O$ / $\rm CaCl_2 \cdot \rm 2H_2O$ or $\rm MgCl_2$ / $\rm Mg(NO_3)_2 \cdot \rm 6H_2O$}  at low temperatures \blue{of 250 -- 500 $^\circ$C} yields not only double perovskite-type $\rm Ca_2MgTeO_6$ and ilmenite-type $\rm Mg_3TeO_6$ polymorphs that \red{were conventionally} only attainable at high temperatures \red{and/}or {\it high pressures} \red{of \blue {>1100$^\circ$C} and \blue {>10 GPa} respectively},\cite {shan2006, selb2019crystal} but also new metastable phases with unique nanocrystal morphology. \red{Despite their sub-optimal pellet density, e}lectrochemical impedance tests reveal \red{$\rm Mg_2CaTeO_6$ and the} high-pressure $\rm Mg_3TeO_6$ polymorph \red{materials} \blue {to} show remarkable $\rm Mg^{2+}$ conductivity at room temperature \red{(}with an activation energy of 0.47 eV noted for the high-pressure $\rm Mg_3TeO_6$ polymorph\red{)}, \red{thus} enlisting a new oxide class of $\rm Mg^{2+}$ conductors. 
\red{Thus, the crowning achievement of our study is successfully} extend\red{ing} the solid-state metathesis reaction to the synthesis of a diverse class of new $\rm Ca$- and $\rm Mg$-containing \blue {layered-} and perovskite-based \red{high-entropy} multifunctional \blue {energy} materials.

\newpage

\section{\label{Section: Results} Results}
 
\magenta {Magnesium- and calcium-based materials were synthesised via solid-state metathesis reactions, utilising target precursors (\textbf {Supplementary Figures 2--6}) with Mg and Ca salts (\textbf {Supplementary Figures 7--12}), with the reactions performed at temperatures below 500 $^\circ$C, as detailed in the \textbf{\nameref{Section: Methods}} section.} The composition of the attained materials were ascertained via energy-dispersive X-ray spectroscopy \magenta {(\textbf {Supplementary Figures 13--40}) or inductively-coupled plasma mass spectrometry (\textbf{Supplementary Table 1})}. Moreover, the purity and crystallinity of the samples prepared via metathesis reactions were verified using X-ray diffraction (XRD) analyses (\textbf{Supplementary Figures 7--12}), as described in the \textbf{\nameref{Section: Methods}} section. The XRD patterns predominantly displayed broad and asymmetric Bragg peaks, a characteristic feature of materials synthesised via metathesis routes. This broadening presents challenges in obtaining detailed atomic structural information, likely due to the reduced crystallinity and inherent disorder of the resulting materials. 
To glean insights into the structure of the materials obtained through solid-state metathetical reactions, particularly regarding the atomic arrangement, we conducted atomic-resolution imaging of the acquired materials along various zone axes. Details regarding sample preparation, measurement protocols, and considerations are elucidated in the \textbf{\nameref{Section: Methods}} section. 

\begin{figure*}[!b]
 \centering
 \includegraphics[width=0.4\columnwidth]{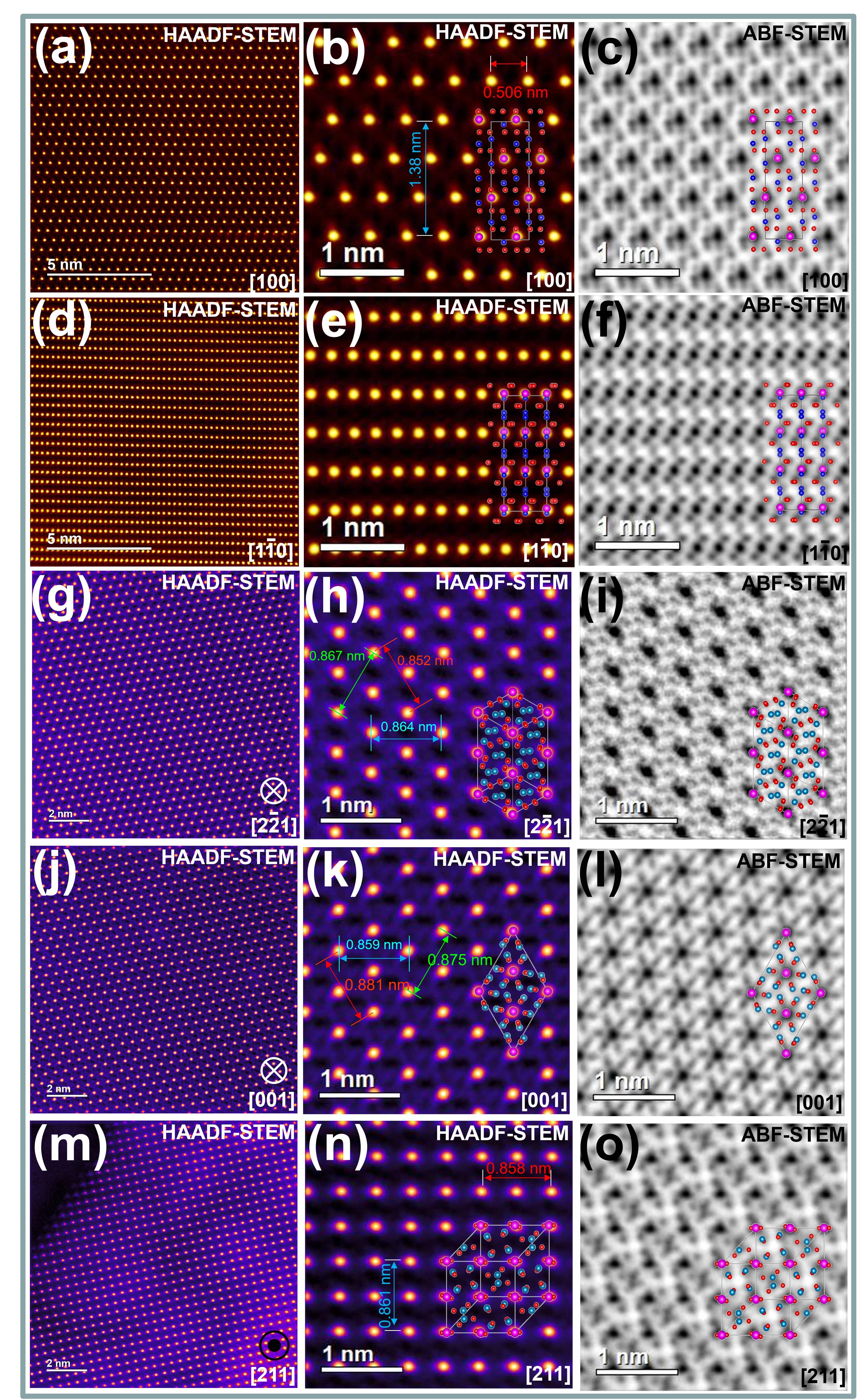}
 \caption{\textbf {: \magenta {Visualisation of the crystal structures of high-pressure, high-temperature $\rm Mg_3TeO_6$ (ilmenite-type) polymorph (prepared via solid-state metathesis reaction of layered $\rm Na_4MgTeO_6$ precursor with anhydrous $\rm MgCl_2$ at 400 $^\circ$C for 99 hours in air. A two-fold molar excess of anhydrous $\rm MgCl_2$ was used.), and ambient-pressure, high-temperature $\rm Mg_3TeO_6$ (corundum-type) polymorph (prepared via solid-state metathesis reaction of $\rm Li_4MgTeO_6$ precursor with anhydrous $\rm MgCl_2$ at 400 $^\circ$C) using high-resolution scanning transmission electron microscopy (STEM). To be visually distinguishable, $\rm Mg$ atoms are shown in (dark/light) blue whilst $\rm Te$ and $\rm O$ atoms are shown in pink and red, respectively.}} \textbf {(a)} High-angle annular dark-field (HAADF)-STEM image of high-pressure, high-temperature $\rm Mg_3TeO_6$ polymorph crystallite taken along the [100] zone axis. \textbf {(b)} Enlarged HAADF-STEM image and \textbf {(c)} Corresponding annular bright-field (ABF)-STEM image. \textbf {(d)} HAADF-STEM image of high-pressure, high-temperature $\rm Mg_3TeO_6$ crystallite taken along the [1$\overline{1}$0] zone axis. \textbf {(e)} Enlarged HAADF-STEM image and \textbf {(f)} Corresponding ABF-STEM image. An atomistic model of the average structure of high-pressure, high-temperature $\rm Mg_3TeO_6$ acquired based on STEM analyses along the [100] and [1$\overline{1}$0] zone axes has been embedded on the STEM images. \textbf {(g)} HAADF-STEM image of ambient-pressure, high-temperature $\rm Mg_3TeO_6$ polymorph crystallite taken along the [221] zone axis and \textbf {(h)} Corresponding ABF-STEM image. \textbf {(i)} Enlarged ABF-STEM image taken along the [221] zone axis. \textbf {(j)} HAADF-STEM image of ambient-pressure, high-temperature $\rm Mg_3TeO_6$ polymorph crystallite taken along the [001] zone axis and \textbf {(k)} Corresponding ABF-STEM image. \textbf {(l)} Enlarged ABF-STEM image taken along the [001] zone axis. \textbf {(m)} HAADF-STEM image of ambient-pressure, high-temperature $\rm Mg_3TeO_6$ polymorph crystallite taken along the [211] zone axis and \textbf {(n)} Corresponding ABF-STEM image. \textbf {(o)} Enlarged ABF-STEM image taken along the [211] zone axis. An atomistic model of the average structure of ambient-pressure, high-temperature $\rm Mg_3TeO_6$ polymorph acquired based on STEM analyses along the [221], [001] and [211] zone axes has been embedded on the STEM images.}
 \label{Figure1}
\end{figure*}

\begin{figure*}[!b]
 \centering
 \includegraphics[width=0.73\columnwidth]{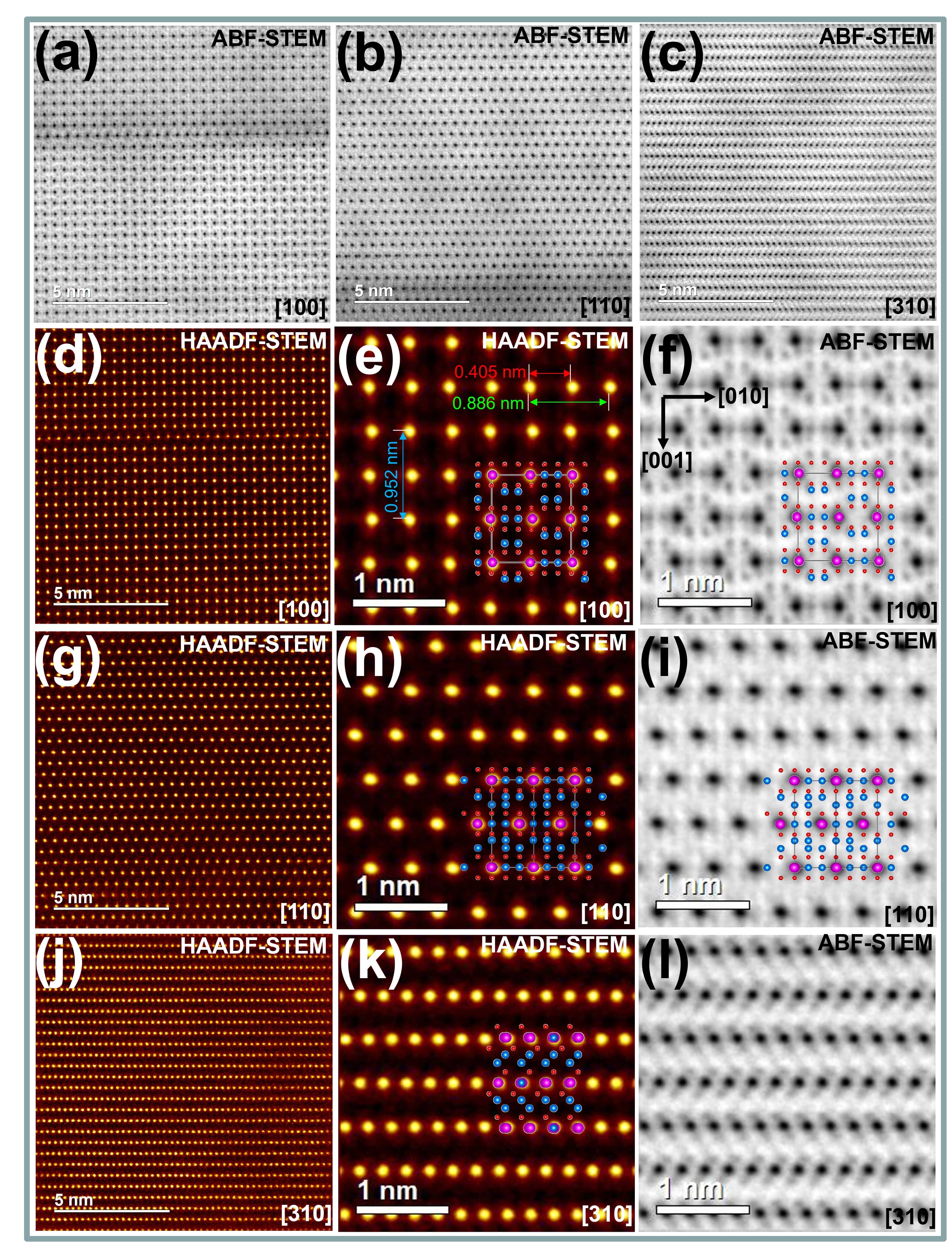}
 \caption{\textbf {: \magenta {Visualisation of the crystal structure of vacancy-type $\rm Mg_3TeO_6$ (prepared via solid-state metathesis reaction of honeycomb-layered $\rm Na_2Mg_2TeO_6$ precursor with $\rm Mg(NO_3)_2$·$\rm 6H_2O$ at 250 $^\circ$C for 48 hours in $\rm N_2$ under ambient pressure) using high-resolution STEM. To be visually distinguishable, $\rm Mg$ atoms are shown in (dark/light) blue whilst $\rm Te$ and $\rm O$ atoms are shown in pink and red, respectively.}} ABF-STEM image of vacancy-type $\rm Mg_3TeO_6$ (low-temperature polymorph) crystallite taken along \textbf {(a)} [100], \textbf {(b)} [110] and \textbf {(c)} [310] zone axes.  
 \textbf {(d)} HAADF-STEM image of $\rm Mg_3TeO_6$ crystallite taken along the [100] zone axis. \textbf {(e)} Enlarged HAADF-STEM image and \textbf {(f)} Corresponding ABF-STEM image. \textbf {(g)} HAADF-STEM image of $\rm Mg_3TeO_6$ crystallite taken along the [110] zone axis. \textbf {(h)} Enlarged HAADF-STEM image and \textbf {(i)} Corresponding ABF-STEM image. \textbf {(j)} HAADF-STEM image of $\rm Mg_3TeO_6$ crystallite taken along the [310] zone axis. \textbf {(k)} Enlarged HAADF-STEM image and (l) Corresponding ABF-STEM image. An atomistic model of the average structure of $\rm Mg_3TeO_6$ acquired based on STEM analyses along the [100], [110] and [310] zone axes has been embedded on the STEM images.}
 \label{Figure 2}
\end{figure*}

\begin{figure*}[!b]
 \centering
 \includegraphics[width=0.5\columnwidth]{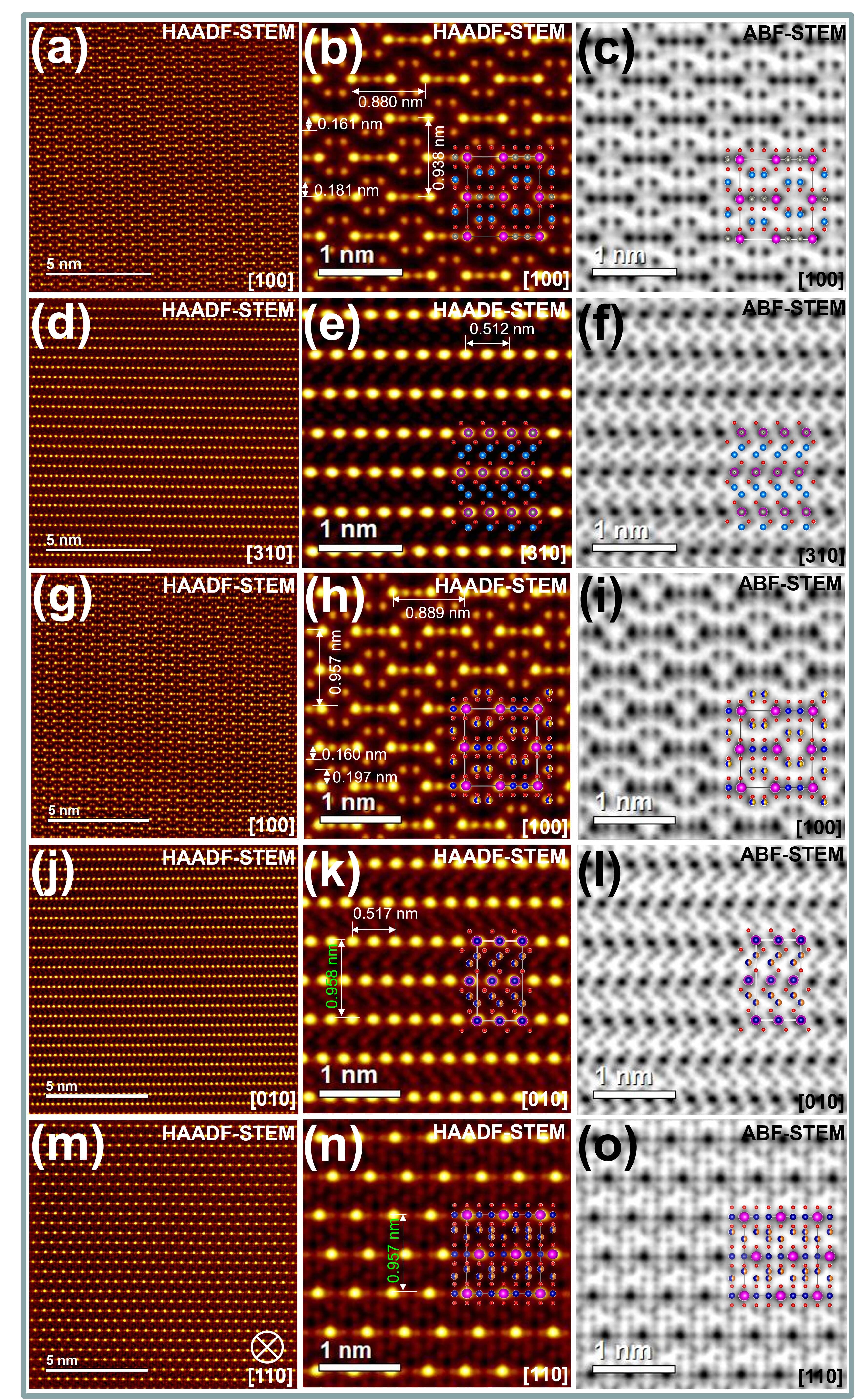}
 \caption{\textbf {: \magenta {Visualisation of the crystal structures of vacancy-type $\rm MgZn_2TeO_6$ and $\rm MgCo_2TeO_6$ (prepared via solid-state metathesis reaction of honeycomb-layered $\rm Na_2Zn_2TeO_6$ and $\rm Na_2Co_2TeO_6$ precursors with $\rm Mg(NO_3)_2$·$\rm 6H_2O$ at 250 $^\circ$C for 48 hours in $\rm N_2$ under ambient pressure) using high-resolution STEM.}} \textbf {(a)} HAADF-STEM image of vacancy-type $\rm MgZn_2TeO_6$ crystallite taken along the [100] zone axis. \textbf {(b)} Enlarged HAADF-STEM image and \textbf {(c)} Corresponding ABF-STEM image. \textbf {(d)} HAADF-STEM image of $\rm MgZn_2TeO_6$ crystallite taken along the [310] zone axis. \textbf {(e)} Enlarged HAADF-STEM image and \textbf {(f)} Corresponding ABF-STEM image. An atomistic model of the average structure of vacancy-type $\rm MgZn_2TeO_6$ acquired based on STEM analyses along the [100] and [310] zone axes has been embedded on the STEM images. To be visually distinguishable, $\rm Mg$ atoms are shown in (light) blue whilst $\rm Te$, $\rm Zn$ and $\rm O$ atoms are shown in pink, grey and red, respectively. \textbf {(g)} HAADF-STEM image of vacancy-type $\rm MgCo_2TeO_6$ crystallite taken along the [100] zone axis. \textbf {(h)} Enlarged HAADF-STEM image and \textbf {(i)} Corresponding ABF-STEM image. \textbf {(j)} HAADF-STEM image of $\rm MgCo_2TeO_6$ crystallite taken along the [010] zone axis. \textbf {(k)} Enlarged HAADF-STEM image and \textbf {(l)} Corresponding ABF-STEM image. \textbf {(m)} HAADF-STEM image of $\rm MgCo_2TeO_6$ crystallite taken along the [110] zone axis. \textbf {(n)} Enlarged HAADF-STEM image and \textbf {(o)} Corresponding ABF-STEM image. Atomistic models of the average structure of $\rm MgCo_2TeO_6$ acquired based on STEM analyses along the [100], [010] and [110] zone axes have been embedded on the STEM images. To be visually distinguishable, $\rm Mg$ atoms are shown in (light) brown whilst $\rm Te$ and $\rm Co$ atoms are shown in pink and blue, respectively. Oxygen atoms are shown in red.}
 \label{Figure3}
\end{figure*}

\begin{figure*}[!t]
 \centering
 \includegraphics[width=0.75\columnwidth]{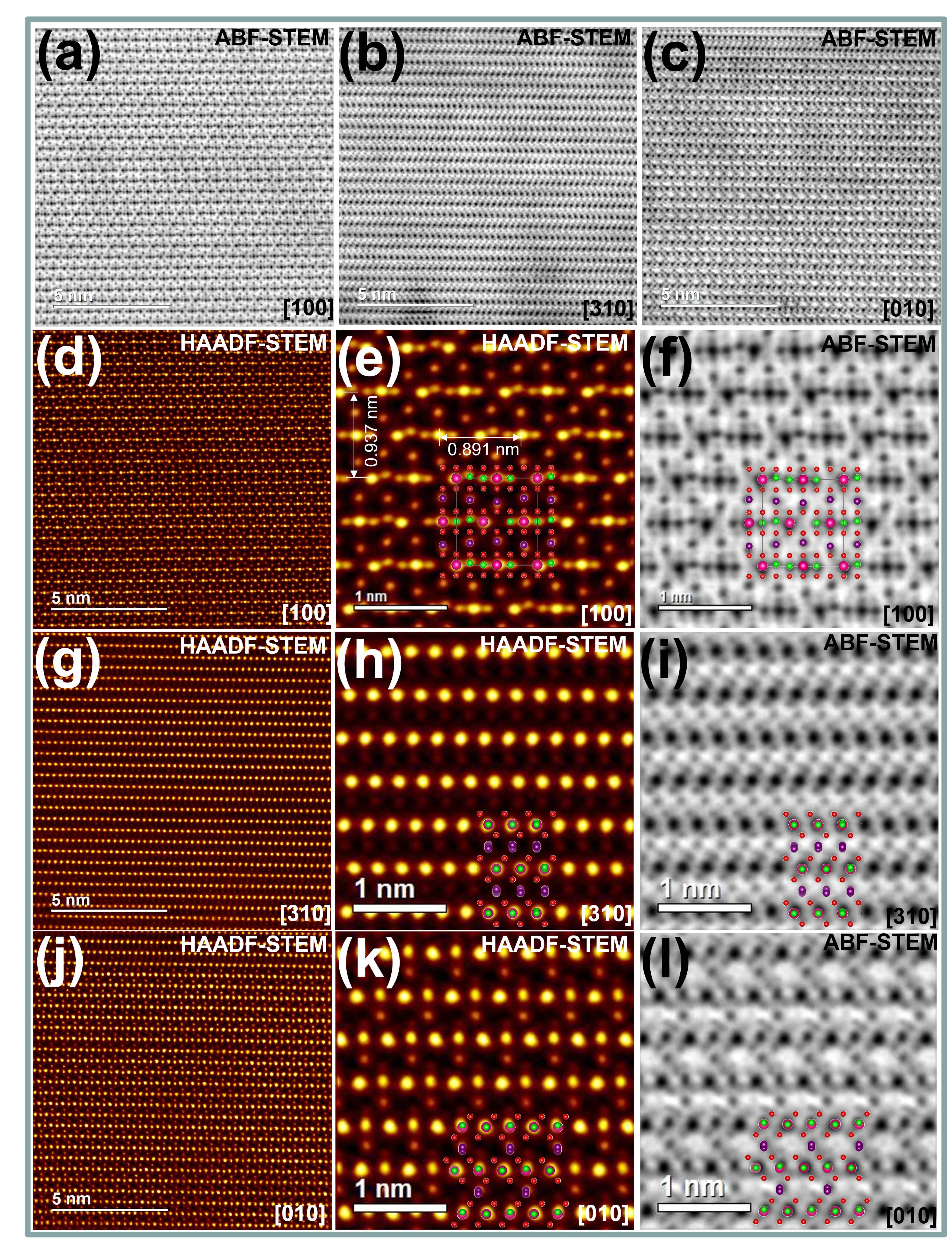}
 \caption{\textbf {: \magenta{Visualisation of the crystal structure of $\rm MgNi_2TeO_6$ (prepared using solid-state metathesis reaction of honeycomb-layered $\rm Na_2Ni_2TeO_6$ precursor with anhydrous $\rm MgCl_2$ at 400 $^\circ$C) using high-resolution STEM. To be visually distinguishable, $\rm Mg$ atoms are shown in purple whilst $\rm Ni$, $\rm Te$ and $\rm O$ atoms are shown in green, pink and red, respectively.}} \textbf {(a, b, c)} ABF-STEM image of $\rm MgNi_2TeO_6$ nanocrystallite taken along [100], [310] and [010] zone axes. \textbf {(d)} HAADF-STEM image of $\rm MgNi_2TeO_6$ nanocrystallite taken along [100] zone axis. \textbf {(e)} Enlarged HAADF-STEM image and \textbf {(f)} Corresponding ABF-STEM image.   
 \textbf {(g)} HAADF-STEM image of $\rm MgNi_2TeO_6$ nanocrystallite taken along [310] zone axis. \textbf {(h)} Enlarged HAADF-STEM image and \textbf {(i)} Corresponding ABF-STEM. \textbf {(j)} HAADF-STEM image of $\rm MgNi_2TeO_6$ crystallite taken along the [010] zone axis. \textbf {(k)} Enlarged HAADF-STEM image and \textbf {(l)} Corresponding ABF-STEM image. An atomistic model of the average structure of $\rm MgNi_2TeO_6$ acquired based on STEM analyses along the [100], [310] and [010] zone axes has been embedded on the STEM images. 
}
 \label{Figure 5}
\end{figure*}

\begin{figure*}[!t]
 \centering
 \includegraphics[width=0.69\columnwidth]{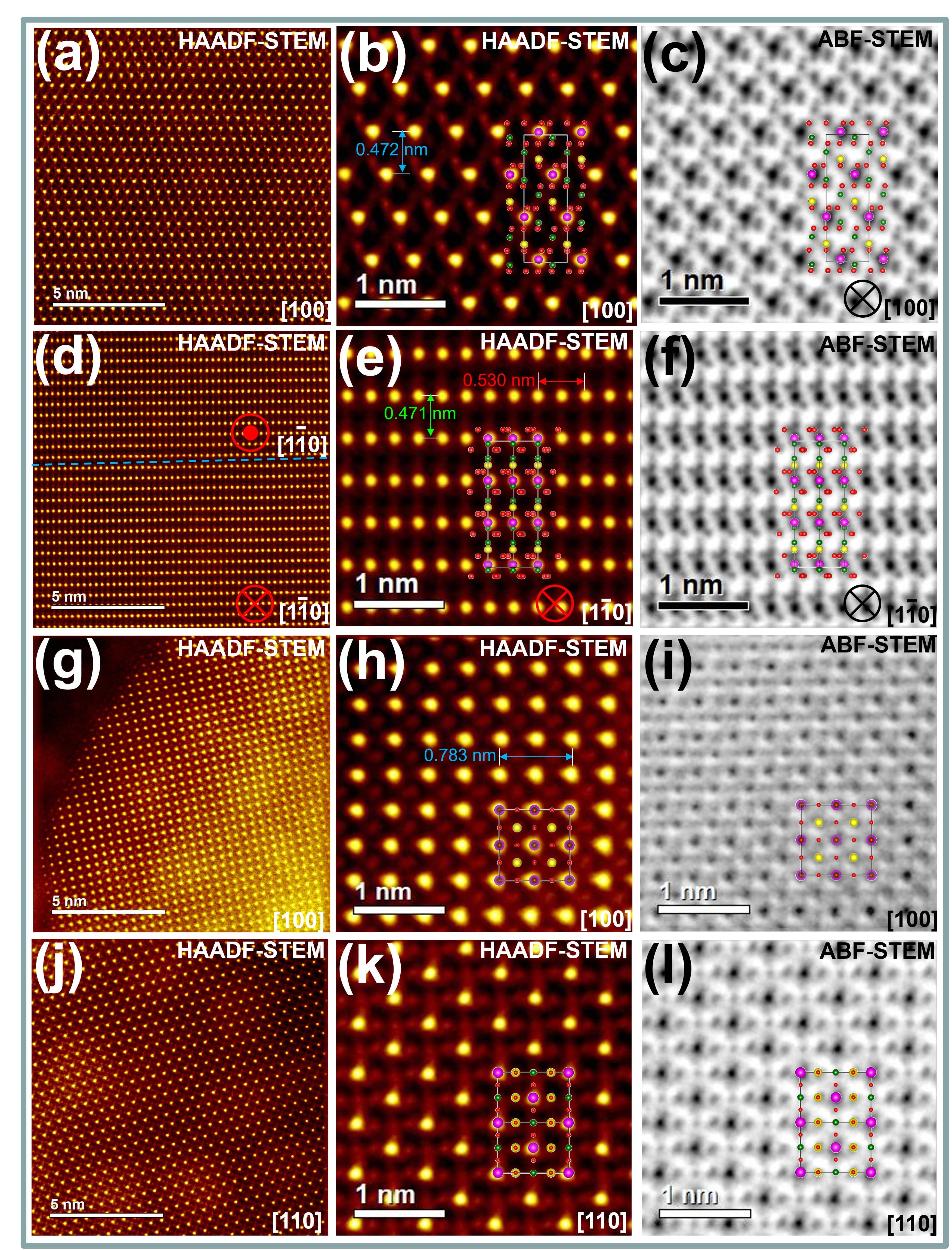}
 \caption{\textbf {: Visualisation of the crystal structure of $\rm CaMg_2TeO_6$ and $\rm Ca_2MgTeO_6$ \magenta {(prepared using solid-state metathesis reaction of honeycomb layered $\rm Na_2Mg_2TeO_6$ and $\rm Na_4MgTeO_6$ precursors with $\rm Ca(NO_3)_2$·$\rm 4H_2O$ at 500 $^\circ$C) for 99 hours in $\rm N_2$} using high-resolution STEM. \red{To be visually distinguishable, $\rm Ca$ atoms are shown in yellow whilst $\rm Te$, $\rm Mg$ and $\rm O$ atoms are shown in pink, green and red, respectively.}} \textbf {(a)} HAADF-STEM image of $\rm CaMg_2TeO_6$ nanocrystallite taken along the [100] zone axis. \textbf {(b)} Enlarged HAADF-STEM image and \textbf {(c)} Corresponding ABF-STEM image showing the aperiodic structural framework.     
 \textbf {(d)} HAADF-STEM image of $\rm Ca_2MgTeO_6$ nanocrystallite taken along the [1$\overline{1}$0] zone axis. \textbf {(e)} Enlarged HAADF-STEM image and \textbf {(f)} Corresponding ABF-STEM. An atomistic model of the average structure of $\rm CaMg_2TeO_6$ acquired based on STEM analyses along the [100] and [1$\overline{1}$0] zone axes has been embedded on the STEM images.\textbf {(g)} HAADF-STEM image of $\rm Ca_2MgTeO_6$ crystallite taken along the [100] zone axis. \textbf {(h)} Enlarged HAADF-STEM image and \textbf {(i)} Corresponding ABF-STEM. \textbf {(j)} HAADF-STEM image of $\rm Ca_2MgTeO_6$ crystallite taken along the [110] zone axis. \textbf {(k)} Enlarged HAADF-STEM image and \textbf {(l)} Corresponding ABF-STEM. An atomistic model of the average structure of $\rm Ca_2MgTeO_6$ acquired based on STEM analyses along the [100] and [110] zone axes has been embedded on the STEM images.}
 \label{Figure 6}
\end{figure*}

\begin{figure*}[!t]
 \centering
 \includegraphics[width=0.85\columnwidth]{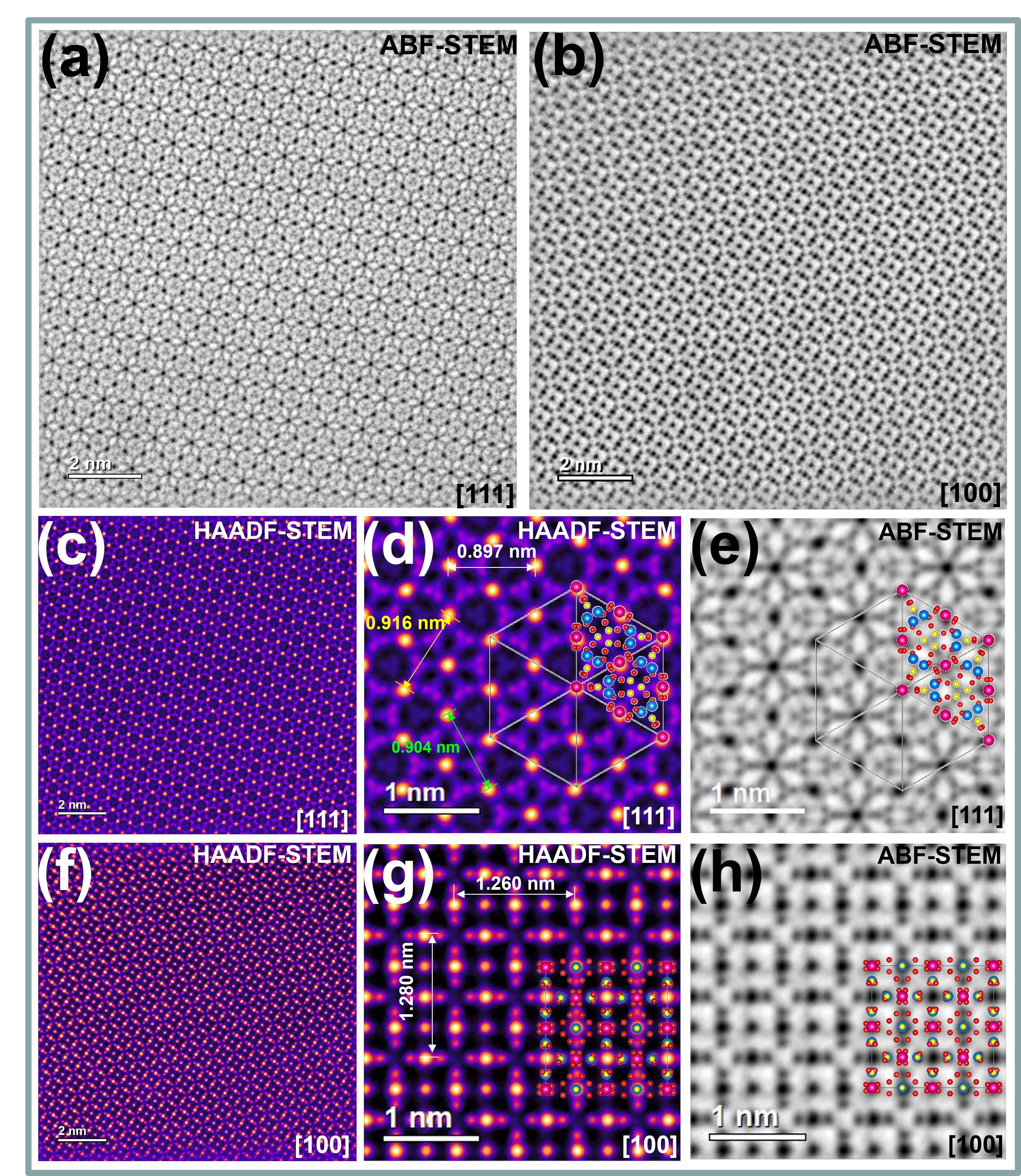}
 \caption{\textbf {: \magenta{Visualisation of the crystal structure of $\rm CaZn_2TeO_6$ (prepared via solid-state metathesis reaction of honeycomb-layered $\rm Na_2Zn_2TeO_6$ precursor with anhydrous $\rm Ca(NO_3)_2$·$\rm 4H_2O$ at 500 $^\circ$C for 99 hours in $\rm N_2$) using high-resolution STEM. To be visually distinguishable, $\rm Ca$ atoms are shown in yellow whilst $\rm Te$, $\rm Zn$ and $\rm O$ atoms are shown in pink, (light)blue and red, respectively.}} ABF-STEM image of $\rm CaZn_2TeO_6$ nanocrystallite taken along the \textbf {(a)} [111] and \textbf {(b)} [100] zone axes. \textbf {(c)} HAADF-STEM image of $\rm CaZn_2TeO_6$ nanocrystallite taken along the [111] zone axis. \textbf {(d)} Enlarged HAADF-STEM image and \textbf {(e)} Corresponding ABF-STEM image. \textbf {(f)} HAADF-STEM image of $\rm CaZn_2TeO_6$ nanocrystallite taken along the [100] zone axis. \textbf {(g)} Enlarged HAADF-STEM image and \textbf {(h)} Corresponding ABF-STEM image. An atomistic model of the average structure of $\rm CaZn_2TeO_6$ acquired based on STEM analyses along the [111] and [100] zone axes has been embedded on the STEM images.}
 \label{Figure 6_2}
\end{figure*}

\begin{figure*}[!t]
 \centering
 \includegraphics[width=0.72\columnwidth]{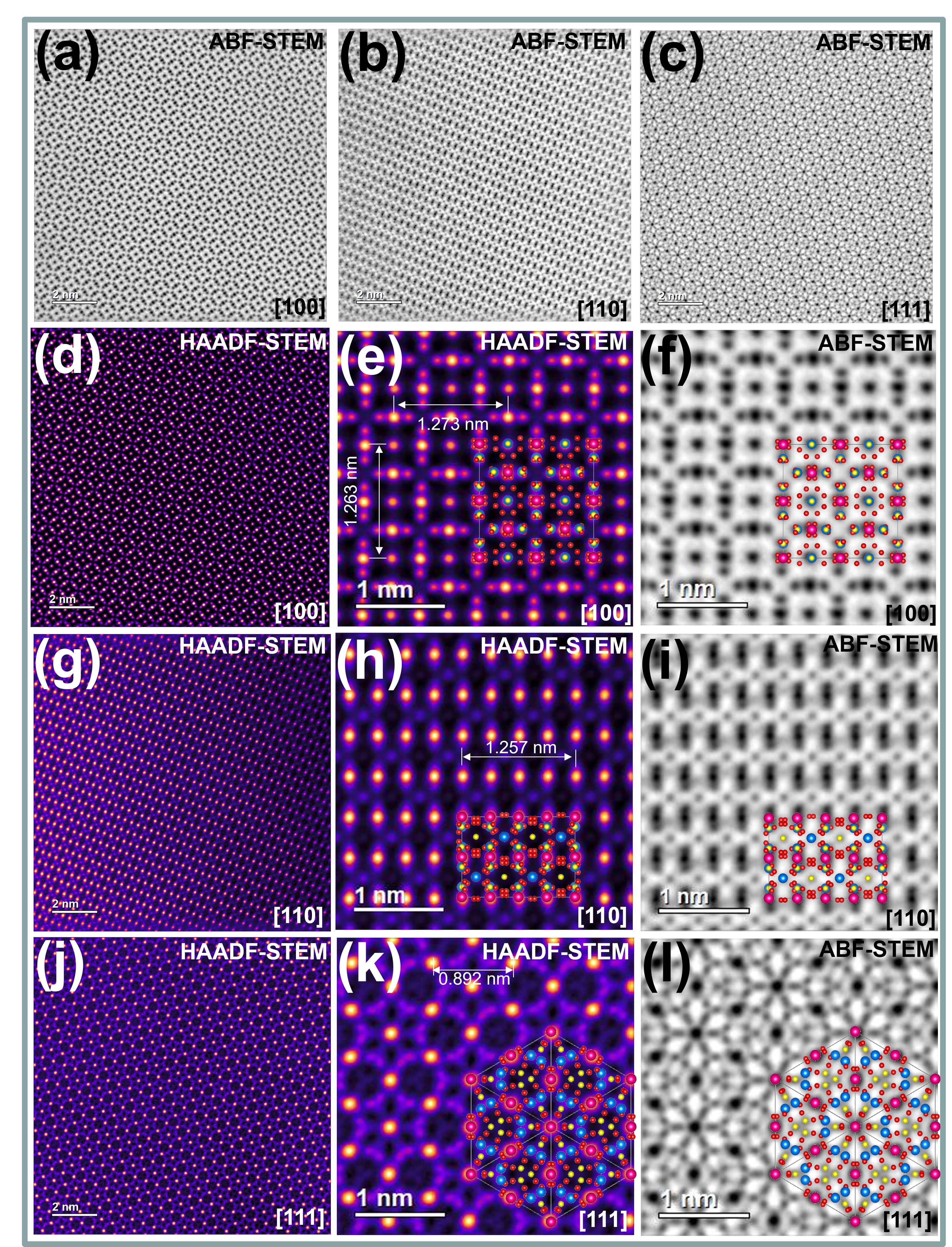}
 \caption{\textbf {: \magenta {Visualisation of the crystal structure of yafsoanite-type $\rm Ca_{1.5}Zn_{1.5}TeO_6$ (prepared via metathesis reaction of honeycomb layered $\rm Na_3Zn_{1.5}TeO_6$ precursor with $\rm Ca(NO_3)_2$·$\rm 4H_2O$ at 500 $^\circ$C for 99 hours in nitrogen atmosphere) using high-resolution STEM. To be visually distinguishable, $\rm Ca$ atoms in $\rm Ca_{1.5}Zn_{1.5}TeO_6$ are shown in yellow whilst $\rm Te$, $\rm Zn$ and $\rm O$ atoms are shown in pink, (light) blue and red, respectively.}} \textbf {(a, b, c)} ABF-STEM image of $\rm Ca_{1.5}Zn_{1.5}TeO_6$ nanocrystallite taken along [100], [110] and [111] zone axes. \textbf {(d)} HAADF-STEM image of $\rm Ca_{1.5}Zn_{1.5}TeO_6$ crystallite taken along [100] zone axis. \textbf {(e)} Enlarged HAADF-STEM image and \textbf {(f)} Corresponding ABF-STEM image. \textbf {(g)} HAADF-STEM image of $\rm Ca_{1.5}Zn_{1.5}TeO_6$ crystallite taken along [110] zone axis. \textbf {(h)} Enlarged HAADF-STEM image and \textbf {(i)} Corresponding ABF-STEM image. \textbf {(j)} HAADF-STEM image of $\rm Ca_{1.5}Zn_{1.5}TeO_6$ crystallite taken along [111] zone axis. \textbf {(k)} Enlarged HAADF-STEM image and \textbf {(l)} Corresponding ABF-STEM image. An atomistic model of the garnet structure of $\rm Ca_{1.5}Zn_{1.5}TeO_6$ acquired based on STEM analyses along the [100], [110] and [111] zone axes has been embedded on the STEM images.}
 \label{Figure 7}
\end{figure*}

\begin{figure*}[!t]
 \centering
 \includegraphics[width=0.7\columnwidth]{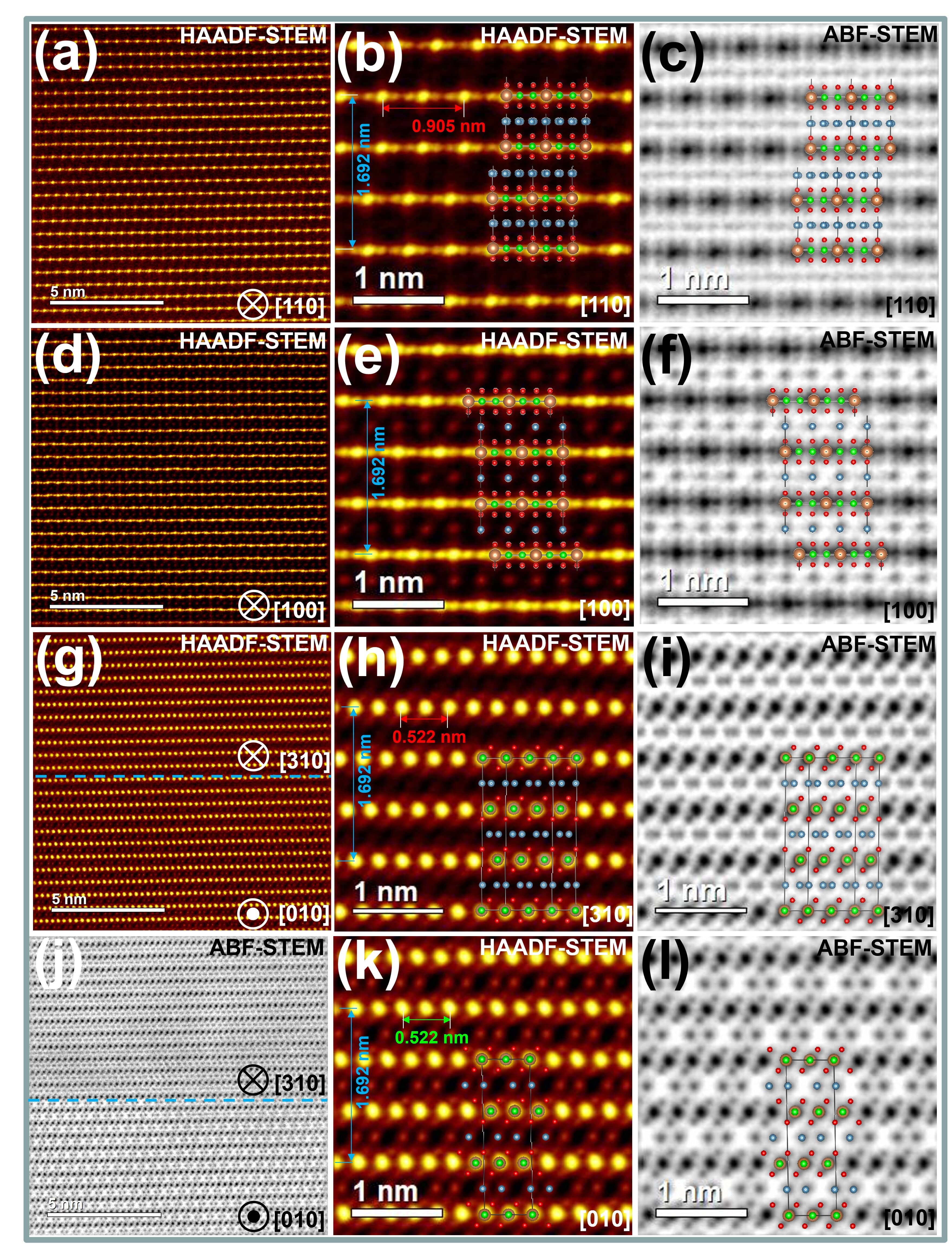}
 \caption{\textbf {: Visualisation of the crystal structure of honeycomb-layered $\rm Ca_{1.5}Ni_2SbO_6$ \magenta {(prepared using solid-state metathesis reaction of honeycomb layered $\rm Na_3Ni_2SbO_6$ precursor with $\rm Ca(NO_3)_2$·$\rm 4H_2O$ at 500 $^\circ$C for 99 hours in nitrogen)} using high-resolution STEM. \red{To be visually distinguishable, $\rm Ca$ atoms are shown in (light) blue whilst $\rm Sb$, $\rm Ni$ and $\rm O$ atoms are shown in yellow, green and red, respectively.}} \textbf {(a)} HAADF-STEM image of $\rm Ca_{1.5}Ni_2SbO_6$ nanocrystallite taken along the [110] zone axis. \textbf {(b)} Enlarged HAADF-STEM image and \textbf {(c)} Corresponding ABF-STEM image showing the aperiodic structural framework. An atomistic model of the average structure of $\rm Ca_{1.5}Ni_2SbO_6$ acquired based on STEM analyses along the [110] zone axis has been embedded on the STEM image.   
 \textbf {(d)} HAADF-STEM image of $\rm Ca_{1.5}Ni_2SbO_6$ nanocrystallite taken along the [100] zone axis. \textbf {(e)} Enlarged HAADF-STEM image and \textbf {(f)} Corresponding ABF-STEM. \textbf {(g)} HAADF-STEM image of $\rm Ca_{1.5}Ni_2SbO_6$ crystallite taken along domains with a juxtaposition of [310] and [010] zone axes. \textbf {(h)} Enlarged HAADF-STEM image taken along the [310] direction and \textbf {(i)} Corresponding ABF-STEM. \textbf {(j)} Low-magnification ABF-STEM image of $\rm Ca_{1.5}Ni_2SbO_6$ crystallite taken along domains with a juxtaposition of [310] and [010] zone axes. \textbf {(k)} Enlarged HAADF-STEM image taken along the [010] direction and \textbf {(l)} Corresponding ABF-STEM.}
 \label{Figure 8}
\end{figure*}

\textbf{Figures {\ref {Figure1}a,b,c,d,e,f}} present high-resolution scanning transmission electron microscopy (STEM) images of $\rm Mg_3TeO_6$, obtained via a solid-state metathetic reaction involving $\rm Mg$ halide 
salt with layered $\rm Na_4MgTeO_6$ at 400 $^\circ$C for 99 hours in air (hypothetically, $\rm Na_4MgTeO_6$ + $\rm 2MgCl_2 \ce{->} \rm Mg_3TeO_6$ + $\rm 4NaCl$). The determined lattice parameters ($a$ = 5.06 \AA, $c$ = 13.8 \AA) closely align with those of the ilmenite-ype $\rm Mg_3TeO_6$ polymorph,\cite {selb2019crystal} typically achieved under high-pressure (12.5 GPa) and high-temperature (1297 $^\circ$C) conditions, crystallising in the non-centrosymmetric (acentric) $R$3 trigonal space group (hereafter referred to as `high-pressure, high-temperature $\rm Mg_3TeO_6$'). Notably, whilst another $\rm Mg_3TeO_6$ polymorph \cite {bhim2018exploring} can be obtained via solid-state reaction of $\rm Li_4MgTeO_6$ and $\rm Mg$ halide salts (\textbf{Figures {\ref {Figure1}g,h,i,j,k,l,m,n,o}} and \textbf{Supplementary Figure 9}) or using solid-state ceramics route under ambient pressure at high temperatures (\textbf{Supplementary Figure 41}), we obtained a previously unreported $\rm Mg_3TeO_6$ polymorph by modifying the initial precursor from layered $\rm Na_4MgTeO_6$ to the honeycomb-layered $\rm Na_2Mg_2TeO_6$ (\textbf{Supplementary Figure 7}). \textbf{Figures {\ref {Figure 2}a,b,c,d,e,f,g,h,i,j,k,l}} display high-resolution STEM images of this polymorph, acquired through a solid-state metathetic reaction involving $\rm Na_2Mg_2TeO_6$ with \magenta {$\rm Mg(NO_3)_2$·$\rm 6H_2O$} salt at 250 $^\circ$C for 48 hours in $\rm N_2$ \magenta {($\rm Na_2Mg_2TeO_6$ + $\rm Mg(NO_3)_2$·$\rm 6H_2O \ce{->} \rm Mg_3TeO_6$ + $\rm 2NaNO_3$ + $\rm 6H_2O$ $\uparrow$)}. \red{The measured lattice parameters for this new polymorph (\blue {hereafter referred to as} vacancy-type $\rm Mg_3TeO_6$) confirm an orthorhombic lattice ($a$ = 5.12 \AA, $b$ = 8.91 \AA, $c$ = 9.37 \AA), crystallising in the possible space groups: {\it Cmcm}, $C{\rm 222_1}$ or $Cmc{\rm 2_1}$.} Whilst $\rm Na_2Mg_2TeO_6$ exhibits a honeycomb arrangement of $\rm Mg$ around $\rm Te$ atoms along the slabs ({\it i.e.}, 
\red{within} the $ab$ plane) following the sequence \red{(along the [100] or equivalently [010] direction) given by} $\rm -Te-Mg-Mg-Te-Mg-Mg-Te-$,\cite {evstigneeva2011new} the metathetic reaction results in \red{the vacancy type $\rm Mg_3TeO_6$ exhibiting a} 
periodic absence of half of the honeycomb array of $\rm Mg$ \red{(denoted by $\rm X$)} within the slab in the sequence: 
\red{$\rm -Te-Mg-Mg-Te-{\rm X}-{\rm X}-Te-$}. 
\red{In fact, due to the high ionic charge of $+2$ for $\rm Mg^{2+}$ relative to that of $\rm Na^{1+}$, the absent $\rm Mg^{2+}$ ions ($\rm X \equiv X(Mg)$) in \textbf{Figures {\ref {Figure 2}f,i}} result from electrostatic ejection from their resident slab into the inter-slab layer.}

\red{This ejection thesis becomes apparent in the case of $\rm MgZn_2TeO_6$, where the absent slab atoms are expected to differ from the inter-slab atoms. In particular,} 
a different 
solid-state metathetic reaction involving honeycomb-layered $\rm Na_2Zn_2TeO_6$ with $\rm Mg$ salt at 400 $^\circ$C for 99 hours in air ($\rm Na_2Zn_2TeO_6$ + $\rm MgCl_2 \ce{->} \rm MgZn_2TeO_6$ + $\rm 2NaCl$) or $\rm Mg(NO_3)_2$·$\rm 6H_2O$ salt at 250 $^\circ$C for 48 hours in $\rm N_2$ ($\rm Na_2Zn_2TeO_6$ + $\rm Mg(NO_3)_2$·$\rm 6H_2O \ce{->} \rm MgZn_2TeO_6$ + $\rm 2NaNO_3$ + $\rm 6H_2O$ $\uparrow$) was performed, resulting in a structure (vacancy-type $\rm MgZn_2TeO_6$) isotypic to that obtained from the metathetic reaction of $\rm Na_2Mg_2TeO_6$ with $\rm Mg$ salt. High-resolution STEM images in \textbf{Figures {\ref {Figure3}a,b,c,d,e,f}} depict the resulting vacancy-type $\rm MgZn_2TeO_6$, revealing the periodic absence of half of the \red{$\rm Zn$} atoms in the honeycomb array 
within the slab that follows the periodic sequence \red{(along the [100] or equivalently [010] direction)}: 
\red{$\rm -Te-Zn-Zn-Te-X-X-Te-$}. 
The absent $\rm Zn$ atoms \red{($\rm X \equiv X(Zn)$)} translocate between the slabs, \red{in near-equal electrostatic competition with $\rm Mg$ in the inter-slab crystallographic sites,} as evidenced by \red{their indistinguishable high-angle annular dark-field (HAADF) intensities.}
The\red{se} crystallographic sites feature two distinct coordination environments with oxygen ligands: $\rm MgO_6$ / $\rm ZnO_6$ octahedral and $\rm MgO_4$ / $\rm ZnO_4$ tetrahedral sites (\textbf{Supplementary Figures 42 and 43}). The measured lattice parameters for $\rm MgZn_2TeO_6$ confirm an orthorhombic lattice, ($a$ = 5.12 \AA, $b$ = 8.80 \AA, $c$ = 9.38 \AA) akin to that of the vacancy-type $\rm Mg_3TeO_6$ (\textbf{Figures {\ref {Figure 2}a,b,c,d,e,f,g,h,i,j,k,l}}). \magenta {Likewise, the solid-state metathesis reaction of honeycomb-layered $\rm Na_2Co_2TeO_6$ with $\rm Mg$ salt at 250 $^\circ$C yields $\rm MgCo_2TeO_6$ ($\rm Na_2Co_2TeO_6$ + $\rm Mg(NO_3)_2$·$\rm 6H_2O \ce{->} \rm MgCo_2TeO_6$ + $\rm 2NaNO_3$ + $\rm 6H_2O$ $\uparrow$), adopting a vacancy-type structure analogous to $\rm MgZn_2TeO_6$ (\textbf{Figures {\ref {Figure3}g,h,i,j,k,l,m,n,o}})} The solid-state metathetic reaction of honeycomb-layered $\rm Na_2Ni_2TeO_6$ with $\rm Mg$ salt at 250 $^\circ$C yields $\rm MgNi_2TeO_6$ \magenta {($\rm Na_2Ni_2TeO_6$ + $\rm MgCl_2 \ce{->} \rm MgNi_2TeO_6$ + $\rm 2NaCl$)}, displaying a unique atomic arrangement (distinct from the \magenta {$\rm Ni_3TeO_6$-type} $\rm MgNi_2TeO_6$ structure attainable through high-temperature ceramics route (\textbf{Supplementary Figure 41})) as depicted in \textbf{Figures {\ref {Figure 5}a,b,c,d,e,f,g,h,i,j,k,l}}.  
A detailed exposition is provided within the \textbf{\nameref{Section: Methods}} and \textbf{Supplementary Information (Supplementary Tables 2--5)} section on the widespread utility of solid-state metathesis reactions, conducted at temperatures not exceeding 400 $^\circ$C and employing designated precursors with \magenta {$\rm Mg$ salts such as $\rm MgCl_2$, $\rm MgSO_4$, $\rm MgSO_4$·$\rm 7H_2O$, $\rm MgBr_2$·$\rm 6H_2O$, $\rm MgCl_2$·$\rm 6H_2O$, or $\rm Mg(NO_3)_2$·$\rm 6H_2O$.} This versatile approach demonstrates its effectiveness in synthesising various $\rm Mg$-based compounds, encompassing ${\rm Mg_2}M{\rm TeO_6}$ ($M = \rm Mg, Ca, Ni, Cu, Co, Zn$), ${\rm Mg_{1.5}}M_{1.5}{\rm TeO_6}$, ${\rm Mg_{1.5}}M_2{\rm BiO_6}$, ${\rm Mg_{1.5}}M_2{\rm SbO_6}$, ${\rm Mg_{1.5}}M_2{\rm RuO_6}$,  and other derivatives (\textbf{Supplementary Table 1}).

\red{Meanwhile, in \textbf{Figures {\ref {Figure 5}e,f}}, a periodic absence of 1/4 of the honeycomb array of $\rm Ni$ within the slab is evident. The missing $\rm Ni$ atoms relocate between the intra- and inter-slabs, thus coexisting with $\rm Mg$, as indicated by the HAADF intensity. In fact, the inter-slab $\rm Mg$ / $\rm Ni$ atoms shift towards the voids (vacancies) in the intra-slabs, resulting in distorted $\rm MgO_6$ / $\rm NiO_6$ octahedra (\textbf{Supplementary Figure 44}). As a result, there is a one-to-one correlation between the emergent displacement pattern in the inter-slabs and the location on the vacancies in the intra-slabs. \textbf{Figures \ref{Figure1}b,c}, \textbf{Figures \ref{Figure 2}e,f}, \textbf{Figures \ref{Figure3}b,c} and \textbf{Figures \ref{Figure3}h,i} also exhibit this correlation between vacancies and displacements observed in \textbf{Figures {\ref {Figure 5}e,f}}, corresponding to global wave-like or sawtooth-like patterns (akin to a sawtooth function or castle rim function) across the $a, b$-axes, stacked in-phase ($0^{\circ}$) and/or out-of-phase ($180^{\circ}$) along the $c$-axis ({\it i.e.}, across the slabs), \textit{albeit} indiscernible at the local level of the unit cell (\textbf{Supplementary Figure 45}). This kind of emergence of periodicity is antagonistic to concepts such as aperiodic lattices/tilings (\textit{e.g.} the Einstein problem\cite{klaassen2022forcing}), whereby locally identical unit cells leads to emergence of global aperiodicity.} 

In an effort to explore material classes containing calcium, solid-state metathetic reactions were conducted using $\rm Ca$ salt at 500 $^\circ$C (see \textbf{\nameref{Section: Methods}} and \textbf{Supplementary Information} section (\textbf{Supplementary Tables 6 and 7})). The reaction involving honeycomb-layered $\rm Na_2Mg_2TeO_6$ and \magenta {$\rm Ca(NO_3)_2$·$\rm 4H_2O$ ($\rm Na_2Mg_2TeO_6$ + $\rm Ca(NO_3)_2$·$\rm 4H_2O \ce{->} \rm CaMg_2TeO_6$ + $\rm 2NaNO_3$ + $\rm 4H_2O$ $\uparrow$)} at 500 $^\circ$C produces a distinctive framework (\textbf{Figures {\ref {Figure 6}a,b,c,d,e,f}}), where half of the $\rm Mg$ atoms translocate between the (inter/intra-)slabs, resulting in systematically arranged \red{vacancies} within the slabs. This arrangement extends to the positions of $\rm Ca$, $\rm Mg$, and 
\red{vacancies} between these slabs. The crystal structure of $\rm Ca_2MgTeO_6$ (or equivalently $\rm Mg_2CaTeO_6$) mirrors the ilmenite-type $\rm Mg_3TeO_6$, with measured lattice parameters $a$ = 5.30 \AA, $c$ = 14.13 \AA 
 (acentric space group $R$3). $\mathrm{Mg_2CaTeO_6}$ exhibits an expanded unit cell dimension compared to the ilmenite-type $\mathrm{Mg_3TeO_6}$ (\textbf{Supplementary Figure 46}), owing to the larger Shannon-Prewitt ionic radius of $\mathrm{Ca^{2+}}$ in octahedral coordination with oxygen atoms in comparison to $\mathrm{Mg^{2+}}$.\cite {shannon1976revised} Moreover, the $\rm CaO_6$ and $\rm MgO_6$ octahedra exhibit distortion, with each atom deviating significantly towards the \red{vacancies} from the centre. Further, the intricate stacking nature of the slabs is evident in \textbf{ Figure {\ref {Figure 6}d}}, entailing the rotation of Te slabs -- a structural disorder that has also been observed in vacancy-type $\rm MgNi_2TeO_6$ (\textbf{Supplementary Figure 47}), vacancy-type $\rm MgZn_2TeO_6$ (\textbf{Supplementary Figure 48}) and honeycomb-layered $\rm Na_3Ni_2SbO_6$\cite {xiao2020}. 

Furthermore, metathetic reactions, achieved by modifying the initial precursor from $\rm Na_2Mg_2TeO_6$ to $\rm Na_4MgTeO_6$, result in calcium-based double perovskite \blue {and garnet} frameworks. \textbf{Figures {\ref {Figure 6}g,h,i,j,k,l}} show STEM images of the double perovskite ordered $\rm Ca_2MgTeO_6$, obtained through the solid-state metathetic reaction of $\rm Na_4MgTeO_6$ with \magenta {$\rm Ca(NO_3)_2$·$\rm 4H_2O$ ($\rm Na_4MgTeO_6$ + $\rm 2Ca(NO_3)_2$·$\rm 4H_2O \ce{->} \rm Ca_2MgTeO_6$ + $\rm 4NaNO_3$ + $\rm 8H_2O$ $\uparrow$)} at 500 $^\circ$C. $\rm Ca_2MgTeO_6$ crystallises in a cubic lattice with lattice constants $a$ = $b$ = $c$ = 7.83 \AA. However, there is a possibility of lattice distortion leading to the loss of crystal symmetry, as commonly observed in perovskite frameworks (in Glazer notation\cite {woodward1997}: $\rm {\it a}^0{\it b}^0{\it c}^0 \ce{->} \rm {\it a}^{\pm}{\it b}^{\pm}{\it c}^{\pm}$). \magenta {The reaction involving honeycomb-layered $\rm Na_2Zn_2TeO_6$ and $\rm Ca(NO_3)_2$·$\rm 4H_2O$ ($\rm Na_2Zn_2TeO_6$ + $\rm Ca(NO_3)_2$·$\rm 4H_2O \ce{->} \rm CaZn_2TeO_6$ + $\rm 2NaNO_3$ + $\rm 4H_2O$ $\uparrow$) at 500 $^\circ$C yields a garnet framework (\textbf{Figures {\ref {Figure 6_2}a,b,c,d,e,f,g,h}}), isotypic to that of yafsoanite-type $\rm Ca_{1.5}Zn_{1.5}TeO_6$.\cite {jarosch1989yafsoanite} $\rm CaZn_2TeO_6$ crystallises in a cubic lattice ($Ia$$\overline{3}$$d$ space group) with lattice constants $a$ = $b$ = $c$ = 12.7 \AA.} 

The universality of solid-state metathetic reactions extends not only to the synthesis of ${\rm Ca}M_2{\rm TeO_6}$ ($M = \rm Mg, Ca, Ni, Cu, Co, Zn$) and ${\rm Ca}_2M{\rm TeO_6}$ compounds but also to the generation of other new compounds, such as those belonging to classes like ${\rm Ca_{1.5}}M_{1.5}{\rm TeO_6}$, ${\rm Ca_{1.5}}M_2{\rm RuO_6}$, ${\rm Ca}M{\rm SiO_4}$, honeycomb-layered ${\rm Ca_{1.5}}M_2{\rm SbO_6}$, ${\rm Ca_{1.5}}M_2{\rm BiO_6}$,  {\it etc.}, as detailed in the \textbf{\nameref{Section: Methods}} and \textbf{Supplementary Information} section (\textbf{Supplementary Tables 6 and 7}). For instance, solid-state metathesis reaction of honeycomb layered $\rm Na_3Zn_{1.5}TeO_6$
precursor with $\rm Ca(NO_3)_2$·$\rm 4H_2O$ at 500 $^\circ$C for 99 hours in nitrogen atmosphere, yields yafsoanite-type $\rm Ca_{1.5}Zn_{1.5}TeO_6$ (\textbf{Figures {\ref {Figure 7}a,b,c,d,e,f,g,h,i,j,k,l}}) 
Moreover, \textbf{Figures {\ref {Figure 8}a,b,c,d,e,f,g,h,I,j,k,l}} present high-resolution STEM images of honeycomb-layered $\rm Ca_{1.5}Ni_2SbO_6$, obtained via a solid-state metathetic reaction involving \magenta {$\rm Ca(NO_3)_2$·$\rm 4H_2O$} with honeycomb-layered layered $\rm Na_3Ni_2SbO_6$ \magenta {($\rm Na_3Ni_2SbO_6$ + $\rm 1.5Ca(NO_3)_2$·$\rm 4H_2O \ce{->} \rm Ca_{1.5}Ni_2SbO_6$ + $\rm 3NaNO_3$ + $\rm 6H_2O$ $\uparrow$) at 500 $^\circ$C for 99 hours in nitrogen}. The metathesis reaction leads to a topochemical transformation, wherein Na atoms (in octahedral coordination with oxygen) are replaced with Ca atoms (in prismatic coordination with oxygen). In FCC / HCP notation \cite {kanyolo2022advances}, the \red{isomorphic} topochemical transformation can be expressed as: $U_{\rm O}V_{\rm (Ni, Ni, Sb)}W_{\rm O}V_{\rm Na}U_{\rm O}V_{\rm (Ni, Ni, Sb)}W_{\rm O} \,\,etc \ce{->} U_{\rm O}V_{\rm (Ni, Ni, Sb)}W_{\rm O}V_{\rm Ca}U_{\rm O}V_{\rm (Ni, Ni, Sb)}W_{\rm O} \,\,etc.$\red{, both with $\rm Ca$ coordinated in a prismatic manner to $\rm O$ atoms. Here,} \red{the honeycomb arrangement of Ni atoms around Sb is maintained during the topochemical transformation, \textit{modulo} possible finite Burgers vectors and $\rm Ni$ or $\rm Ca$ vacancies. However, due to a transformation from octahedral to prismatic, and the creation of vacancies and other related displacements and distortions in the topochemical transformation, the resultant material is not at all isomorphic to $\rm Na_3Ni_2SbO_6$, as expected above. In fact, the FCC / HCP notation for the obtained polymorph as is evident in \textbf{Figures \ref{Figure 8}e,f} (\textit{modulo} the Burgers vectors,} $\red{\mathbf{b}_-} = [\pm 1/6, -1/2, 0]$ and $\red{\mathbf{b}_+} = [\pm 1/6, +1/2, 0]$ 
\red{implemented respectively by} the shear transformations,
\begin{align}
    S_- = \begin{pmatrix}
1 & 0 & \pm 1/6\\ 
0 & 1 & -1/2\\ 
0 & 0 & 1
\end{pmatrix},\,\,\,S_+ = \begin{pmatrix}
1 & 0 & \pm 1/6\\ 
0 & 1 & \red{+}1/2\\ 
0 & 0 & 1
\end{pmatrix},
\end{align}
\red{and neglecting the $\rm Ca$ vacancies) corresponds instead to:
\begin{multline}\label{FCC_eq}
    W_{\rm O}U_{(\rm Sb, Ni, Ni)}V_{\rm O}U_{\rm Ca}V_{\rm O}U_{(\rm Ni, Sb, Ni)}W_{\rm O}U_{\rm Ca}W_{\rm O}U_{(\rm Ni, Sb, Ni)}V_{\rm O}U_{\rm Ca}V_{\rm O}U_{(\rm Ni, Sb, Ni)}W_{\rm O}U_{\rm Ca}W_{\rm O}U_{(\rm Ni, Ni, Sb)}\\
    V_{\rm O}U_{\rm Ca}V_{\rm O}U_{(\rm Ni, Sb, Ni)}W_{\rm O}U_{\rm Ca}W_{\rm O}U_{(\rm Ni, Sb, Ni)}V_{\rm O}U_{\rm Ca}V_{\rm O}U_{(\rm Ni, Sb, Ni)}W_{\rm O}U_{\rm Ca}W_{\rm O}.
\end{multline}
Unfortunately, the FCC / HCP notation actually fails to appropriately incorporate local information about the location of the vacancies in the \textit{three} hexagonal sub-lattices given by $U_{\rm Ca} = U_{(\rm Ca, Ca, Ca)}$ appearing in eq. (\ref{FCC_eq}), due to the fact that the hexagonal sub-lattice with any number of vacancies departs from an optimised congruent sphere packing.\cite{kanyolo2022advances}} 


\section{\label{Section: Discussion} Discussion}

In employing solid-state metathesis reactions of layered tellurates, we showcase the syntheses of both new and pre-existing $\rm Ca$- and $\rm Mg$-based materials, conventionally achievable not only at high temperatures but also under high pressures. \magenta {For example, the metathetic reaction at temperatures not exceeding 400 $^\circ$C involving layered $\rm Na_4MgTeO_6$ with $\rm MgCl_2$ in ambient pressure ($\rm Na_4MgTeO_6$ + $\rm 2MgCl_2 \ce{->} \rm Mg_3TeO_6$ + $\rm 4NaCl$) or $\rm MgSO_4$ under vacuum ($\rm Na_4MgTeO_6$ + $\rm 2MgSO_4 \ce{->} \rm Mg_3TeO_6$ + $\rm 2Na_2SO_4$) yields the high-pressure, high temperature $\rm Mg_3TeO_6$ (ilmenite-type) polymorph (\textbf{Figures {\ref {Figure1}a,b,c,d,e,f}}), whereas the metathetic reaction involving honeycomb-layered $\rm Li_4MgTeO_6$ with $\rm MgCl_2$ (·$\rm 6H_2O$) or $\rm MgBr_2$·$\rm 6H_2O$ produces the ambient pressure, high-temperature $\rm Mg_3TeO_6$ (corundum-type) polymorph (\textbf{Figures {\ref {Figure1}g,h,i,j,k,l,m,n,o}} and \textbf {Supplementary Figures 7--9, 49 and 50}).} 
Furthermore, the metathetic reaction of honeycomb-layered $\rm Na_2Mg_2TeO_6$ \magenta {with $\rm MgCl_2$ (at 400 $^\circ$C) or $\rm Mg(NO_3)_2$·$\rm 6H_2O$ (at 250 $^\circ$C),} yields a metastable $\rm Mg_3TeO_6$ polymorph (vacancy-type $\rm Mg_3TeO_6$ (\textbf{Figures {\ref {Figure 2}a,b,c,d,e,f,g,h,i,j,k,l}})). The choice of the initial precursor compound ($\rm Li_4MgTeO_6$, $\rm Na_4MgTeO_6$, or $\rm Na_2Mg_2TeO_6$) enables polymorph selectivity of $\rm Mg_3TeO_6$, with the exothermicity of the byproduct halide/nitrate salts, due to their high lattice energy,\cite {parkin1996solid, bonneau1992solid, nartowski1998rapid} attributable to favouring the forward reaction. Moreover, the high lattice energy of the byproduct salt along with the low temperatures employed in the metathesis synthetic route confer an advantage in nanoparticle synthesis (\textbf {Supplementary Figure 51}), in principle by swiftly arresting growth post-nucleation, preventing substantial crystal growth. This contrasts with traditional solid-state synthesis methods, where high reaction temperatures often result in larger particle sizes owing to processes like particle coarsening and Ostwald ripening during crystal growth.\cite {voorhees1985theory} Additionally, we extend the solid-state metathetical approach to synthesise $\rm Ca$-based compounds, such as $\rm CaMgSiO_4$ (\textbf {Supplementary Figure 52}), ilmenite-type $\rm CaMg_2TeO_6$ (or equivalently as $\rm Mg_2CaTeO_6$ (\textbf{ Figures {\ref {Figure 6}a,b,c,d,e,f}}), double-perovskite $\rm Ca_2MgTeO_6$ (\textbf{Figures {\ref {Figure 6}g,h,i,j,k,l}}), and including their derivative compositions (\textbf {Supplementary Tables 6 and 7}). The versatility of the metathetical approach in synthesising a plethora of new $\rm Mg$- and $\rm Ca$-based compounds (some of which are exemplified in \red{equation (\ref{exemplified_eqs})}) is extensively detailed in the \textbf{\nameref{Section: Methods}} and \textbf {Supplementary Information} section.

\begin{figure*}[!t]
 \centering
 \includegraphics[width=0.9\columnwidth]{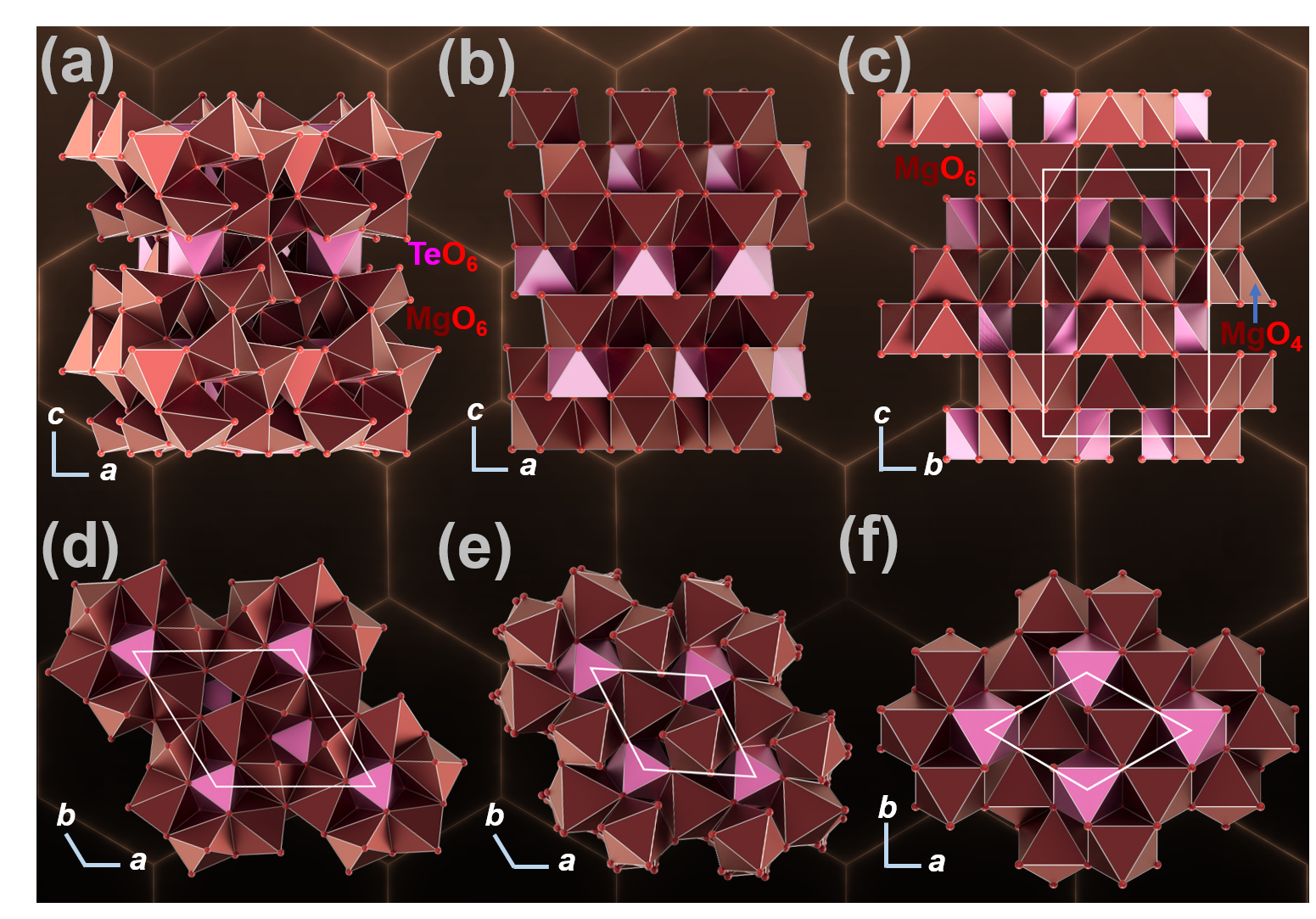}
 \caption{\textbf {: $\rm Mg_3TeO_6$ polymorphs.} \textbf {(a)} Ambient-pressure, high-temperature $\rm Mg_3TeO_6$ (corundum-type) polymorph crystal viewed along the [010] zone axis. \textbf {(b)} High-pressure, high-temperature $\rm Mg_3TeO_6$ (ilmenite-type) polymorph crystal viewed along the [010] zone axis. $\rm Mg$ atom polyhedron ($\rm MgO_6$ or $\rm MgO_4$) is shown in brown, whereas $\rm Te$ atom polyhedron ($\rm TeO_6$) is shown in pink. \textbf {(c)} Vacancy-type (ambient-pressure, low-temperature) $\rm Mg_3TeO_6$ polymorph crystal viewed along the [100] zone axis. \textbf {(d)} Ambient-pressure, high-temperature $\rm Mg_3TeO_6$ polymorph crystal viewed along the [001] zone axis. \textbf {(e)} High-pressure, high-temperature $\rm Mg_3TeO_6$ polymorph crystal viewed along the [001] zone axis. \textbf {(f)} Vacancy-type (ambient-pressure, low-temperature) $\rm Mg_3TeO_6$ polymorph crystal viewed along the [001] zone axis.}
 \label{Figure_5}
\end{figure*}

\textbf{Figure {\ref {Figure_5}}} depicts the crystal structural frameworks of $\rm Mg_3TeO_6$ polymorphs achieved through solid-state metathesis reactions. The high-pressure, high-temperature $\rm Mg_3TeO_6$ (ilmenite-type) polymorph framework (crystallising in the acentric $R$3 space group\cite {selb2019crystal} (\textbf{Figures {\ref {Figure_5}b,e}}) represents a superstructure of the ambient-pressure, high-temperature $\rm Mg_3TeO_6$ (corundum-type) polymorph (\textbf{Figures {\ref {Figure_5}a,d}})), with $\rm Mg^{2+}$ and $\rm Te^{6+}$ cations octahedrally coordinated to oxygen ligands. Additionally, the high-pressure, high-temperature $\rm Mg_3TeO_6$ polymorph exhibits a honeycomb arrangement of $\rm TeO_6$ and $\rm MgO_6$ octahedra (\textbf{Figure {\ref {Figure_5}e}}). One layer is formed by two crystallographically distinct $\rm MgO_6$ octahedra, whilst the second layer consists of the third $\rm MgO_6$ octahedra and the $\rm TeO_6$ octahedra. The alternating sequence of octahedral layers along the $c$-axis, along with two types of layers, results from threefold screw axes (\textbf {Supplementary Figure 53}). The ambient-pressure, high-temperature $\rm Mg_3TeO_6$ (corundum-type) polymorph (crystallising in the centrosymmetric $R$$\overline{3}$ space group\cite {bhim2018exploring}) features centers of inversion symmetry. Structural distinctions arising from the added inversion symmetry centers at the edge centers and corners of the unit cell (viewed along the $c$-axis) are clearly discernible in \textbf{Figures {\ref {Figure_5}d,e}}. In both crystal structures, the $\rm TeO_6$ octahedra are separated and not directly connected. The linkage occurs exclusively via $\rm MgO_6$ octahedra, with six and three edges of the $\rm TeO_6$ octahedra in the ambient-pressure, high-temperature $\rm Mg_3TeO_6$ and the high-pressure, high-temperature $\rm Mg_3TeO_6$ shared with the $\rm MgO_6$ octahedra, respectively (see also \textbf {Supplementary Figure 53}). Vacancy-type $\rm Mg_3TeO_6$ (ambient-pressure, low-temperature) polymorph exhibits a similar structural motif as the high-pressure, high-temperature $\rm Mg_3TeO_6$ polymorph when viewed along the $c$-axis (\textbf{Figures {\ref {Figure_5}c,f}}), featuring distorted $\rm MgO_6$, $\rm MgO_5$ and $\rm MgO_4$ polyhedra coordinated with $\rm TeO_6$ (\textbf{Supplementary Table 8}), resulting in a structure replete with vacancies. Oxide materials featuring five-coordinated $\rm MgO_5$ polyhedral structures are uncommon; therefore, the vacancy-type $\rm Mg_3TeO_6$ (ambient-pressure, low-temperature polymorph) can be regarded as a metastable phase. With regards to structural considerations and the likelihood of obtaining $\rm Mg_3TeO_6$ polymorphs (\textbf {Supplementary Figure 52}), the stability of $\rm Mg_3TeO_6$ polymorphs can be inferred to follow the sequence: ambient-pressure, high-temperature $\rm Mg_3TeO_6$ (corundum-type) > high-pressure, high-temperature $\rm Mg_3TeO_6$ (ilmenite-type) > vacancy-type (ambient-pressure, low-temperature) $\rm Mg_3TeO_6$. \magenta {Structural stability calculations based on density functional theory (\textbf{Supplementary Note 1} and \textbf{Supplementary Table 9}) further corroborate this hypothesis. Moreover, XRD measurements (\textbf{Supplementary Figure 54}) show the vacancy-type $\rm Mg_3TeO_6$ polymorph to transform to the most stable corundum-type $\rm Mg_3TeO_6$ polymorph (ambient pressure, high temperature phase) at high temperatures (800$^{\circ}$C).}

\begin{figure*}[!t]
 \centering
 \includegraphics[width=0.92\columnwidth]{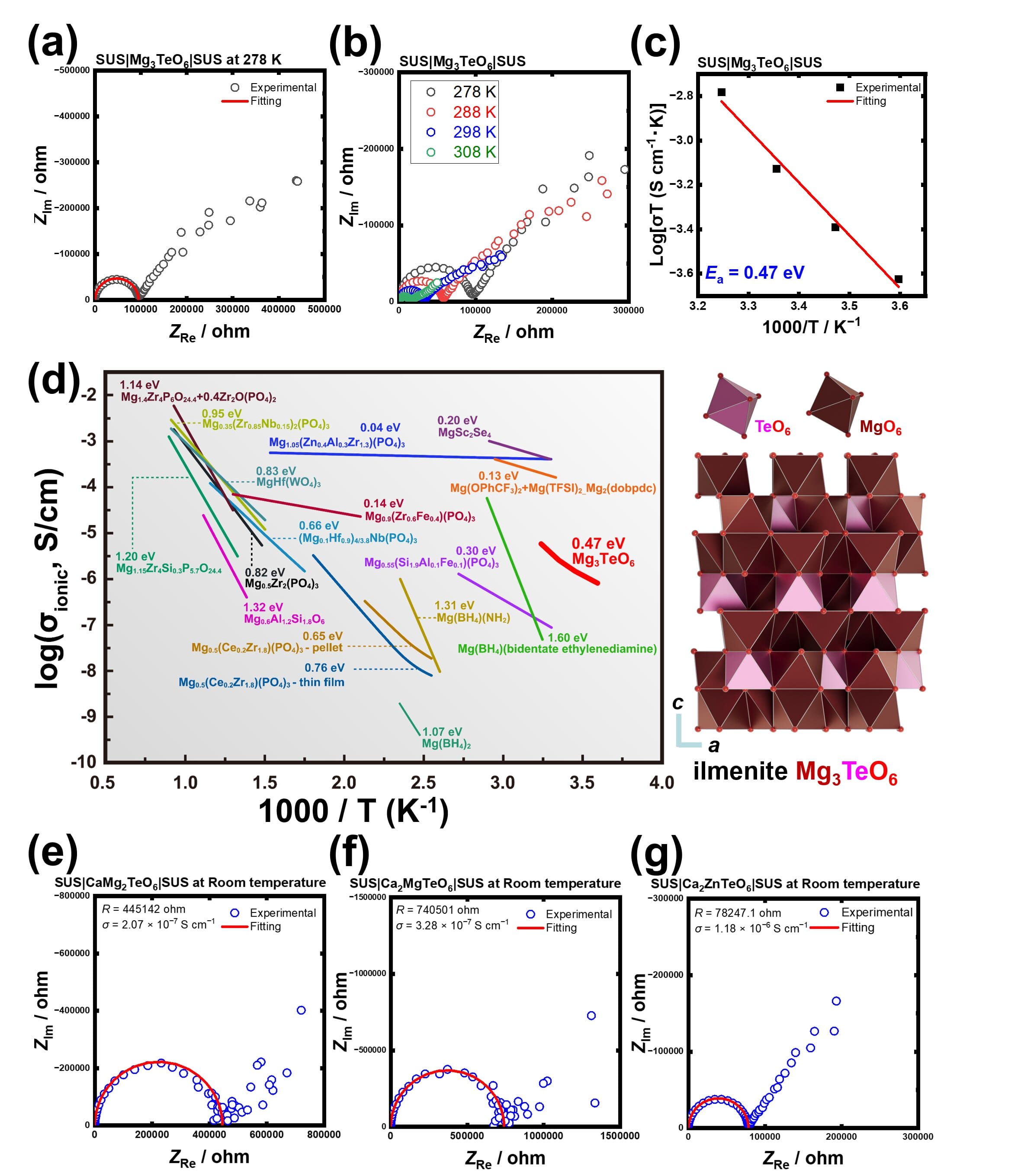}
 \caption{\textbf{: Electrochemical impedance spectroscopy (EIS) measurements of tellurate-based $\rm Mg^{2+}$ and $\rm Ca^{2+}$ conductors.} \textbf {(a)} Nyquist plots taken at 278 K. \textbf {(b)} Nyquist plots taken at various temperatures. \textbf {(c)} Arrhenius plots derived from EIS measurements of ilmenite-type $\rm Mg_3TeO_6$ \magenta {(high pressure, high temperature polymorph)} at various temperatures. \textbf {(d)} Comparative plots of the ionic conductivity values ($\sigma \rm_{ionic} $) against temperature (T) attained in representative Mg-based ionic conductors reported along with the ilmenite-type $\rm Mg_3TeO_6$. Details regarding the measurements and synthesis techniques are provided in the \textbf{\nameref{Section: Methods}} section. \textbf {(e,f,g)} Nyquist impedance plots of $\rm CaMg_2TeO_6$ (or equivalently as $\rm Mg_2CaTeO_6$), $\rm Ca_2MgTeO_6$ and $\rm Ca_2ZnTeO_6$ taken at room temperature.}
 \label{Figure_6}
\end{figure*}

From a technological standpoint, we assessed the viability of employing $\rm Mg_3TeO_6$ polymorphs as $\rm Mg^{2+}$ conductors. \magenta {Attenuated total reflection Fourier-transform infrared ($\rm ATR$-$\rm FTIR$) and nuclear magnetic resonance ($\rm ^1H$ magic-angle spinning ($\rm MAS$) NMR) measurements on the material powder confirmed the absence of $\rm OH^-$ and $\rm H^+$ species (\textbf {Supplementary Figures 55 and 56}), which could otherwise influence the ionic conductivity of the samples.} Whilst the ambient-pressure, high-temperature $\rm Mg_3TeO_6$ (corundum-type) and vacancy-type $\rm Mg_3TeO_6$ (ambient pressure, low-temperature) polymorphs display inferior $\rm Mg^{2+}$ conductivity at room temperature (\textbf {Supplementary Figure 57}), the high-pressure, high-temperature $\rm Mg_3TeO_6$ (ilmenite-type) polymorph demonstrates remarkable ionic conductivity (\textbf{Figures {\ref {Figure_6}a,b}}), despite sub-optimal pellet density. The ionic conductivities of the high-pressure, high-temperature $\rm Mg_3TeO_6$ polymorph at different temperatures were evaluated, as depicted by the Arrhenius plots in \textbf{Figure {\ref {Figure_6}c}}. Details regarding the experimental procedures are outlined in the \textbf{\nameref{Section: Methods}} section. In the temperature range of 5--45 $^{\circ}$C, the high-pressure, high-temperature $\rm Mg_3TeO_6$ polymorph (with a pellet compactness of $\sim$60 \%) exhibited an activation energy of approximately 0.47 eV, calculated by fitting the alternating current data with the Arrhenius equation. The material's ionic conductivity was determined to be $2.5 \times 10^{-6} \,\,\rm S\,cm^{-1}$ at 25 $^{\circ}$C. Bulk ionic conductivities of the high-pressure, high-temperature $\rm Mg_3TeO_6$ polymorph are presented for performance benchmarking against reported $\rm Mg^{2+}$ conductors\cite {jaschin2020, guo2022solid} (\textbf{Figure {\ref {Figure_6}d}}). This figure highlights the potential of the high-pressure, high-temperature $\rm Mg_3TeO_6$ polymorph as a promising solid electrolyte for all-solid-state magnesium batteries. Further improvements in the ionic conductivity can be anticipated, once the pellet densification process is optimised. Furthermore, we explored the possible use of $\rm Ca$-based tellurates as $\rm Ca^{2+}$ conductors, considering the scarcity of oxide compounds with appreciable ionic conductivity at room temperature.\cite {shinde2023} \textbf{Figures {\ref {Figure_6}e,f,g}} depict the electrochemical impedance plots of $\rm CaMg_2TeO_6$ and double perovskites ${\rm Ca_2}M{\rm TeO_6}$ ($M = \rm Mg, Zn $). Although the possibility of mixed $\rm Ca^{2+}$ and $\rm Mg^{2+}$ conductivity cannot be disregarded in $\rm Ca_2MgTeO_6$ and $\rm CaMg_2TeO_6$, double-perovskite $\rm Ca_2ZnTeO_6$, in particular, exhibits an ionic conductivity of $1.18 \times 10^{-6} \,\,\rm S\,cm^{-1}$ at room temperature, primarily arising from $\rm Ca^{2+}$ conduction. This result 
envisages the broad class of $\rm Ca$-based double perovskites (\textit {viz}., ${\rm Ca_2}M{\rm TeO_6}$, ${\rm Ca_{1.5}}M_2{\rm RuO_6}$ and ${\rm Ca_2}M{\rm WO_6}$ ($M = \rm Mg, Zn, Ca$)\cite {burckhardt1982, shan2006, blasse1965new, patwe2006, baglio1969}), ${\rm Ca_{1.5}}M_2{\rm BiO_6}$ and honeycomb-layered ${\rm Ca_{1.5}}M_2{\rm SbO_6}$ (\textbf {Supplementary Tables 6 and 7}), as potential $\rm Ca^{2+}$ conductors in calcium battery solid electrolytes.
\\The additional advantage of the metathesis reactions involving $\rm Li$-, $\rm Na$-, and $\rm K$-based materials with \magenta {$\rm Ca(NO_3)_2$·\\$\rm 4H_2O$/$\rm CaCl_2$·$\rm 2H_2O$ or $\rm MgCl_2$/$\rm Mg(NO_3)_2$·$\rm 6H_2O$/$\rm MgSO_4$} lies in the spontaneous occurrence of the reactions without the need for pellet compactification to attain $\rm Ca$-based and $\rm Mg$-based compounds, as required, for instance, in the preparation of $\rm Ca_{\it x}CoO_2$ ($0.26 < {\it x} \leq0.50$)\cite {cushing1998topotactic}. \magenta {These metathesis reactions initiate spontaneously once the necessary activation energy, $E_{\rm a}$ is reached. For activation facilitated by changes in the lattice internal energy, the necessary activation energy can be understood via the first law of thermodynamics, $dU = TdS - PdV - \mu dN \equiv E_{\rm a}$, where $U = TS - PV$ is the internal energy of the unit cell, $T$ is the temperature, $S$ is the entropy, $P$ is the pressure, $V$ is the volume and the Gibbs-Duhem equation, $\mu dN = -SdT + VdP$ with $\mu$ the chemical potential and $N$ the total number of $A$-based cations and their vacancies in each unit cell. At constant pressure, particle number and heat energy $\delta Q = TdS$, sufficient activation energy can only be obtained by a net-negative change in the volume of the unit cell. Indeed, this change correlates with a change of the coordination of the $A$-cation from prismatic (\textit{e.g.} $\rm Na_2Mg_2TeO_6$) to octahedral (vacancy-type $\rm Mg_3TeO_6$) in the chemical reactions exemplified in eq. (\ref{exemplified_eqs}). On the other hand, vacancy creation and annihilation can be treated as a subset of lattice distortions. Such distortions must increase the entropy of the system ($dS > 0$) injecting disorder into the unit cell. Thus, at constant volume, pressure and temperature, lattice distortions are ideally exothermic ($\delta Q = TdS > 0$), providing additional lattice energy for activation at ambient pressure and considerably lower temperatures. This is the basis of characterising the synthesised materials ridden with such crystalline distortions as high-entropy -- a concept reminiscent of nanostructured high-entropy alloys.\cite{yeh2004nanostructured}} Although \red{a precise analysis of the isobaric, isothermal and mechanistic} pathway is \red{more} intricate and requires computational insights beyond the scope of the present study, the metathesis reactions detailed herein presumably proceed via an ionic route\cite {treece1994metathetical} \red {{\it e.g.}, 
\begin{multline}
    {\rm Na_2}M_2{\rm TeO_6}\,\,(M = {\rm Ni, Co}) + {\rm MgCl_2} \ce{->} {\rm Na_2}^{+}{\it M}_2^{2+}[{\rm TeO_6}]^{6+} + {\rm Mg^{2+}} + {\rm 2Cl^{-}}\\
    \ce{->} {\rm Mg^{2+}}{\it M}_2^{2+}[{\rm TeO_6}]^{6+} + {\rm 2Na^{+}} + {\rm 2Cl^{-}} \ce{->} {\rm Mg}{\it M}_2{\rm TeO_6} + {\rm 2NaCl},
\end{multline}}
forming the target material via atomic rearrangements and no change in the oxidation state of the redox-active transition metal atom constituents. Magnetic susceptibility measurements (\textbf{Supplementary Figure 57}) indicate that the oxidation states of transition metal atoms remain unaltered after the completion of metathesis reactions {\it e.g.}, involving ${\rm Na_2}M_2{\rm TeO_6}$ ($M = \rm Co, Ni$) with $\rm MgCl_2$ to form ${\rm Mg}M_2{\rm TeO_6}$ and $\rm NaCl$ end products. 

\magenta {Finally, direct measurement of the maximum temperatures attained in solid-state metathesis reactions proved challenging due to the lack of a reliable thermocouple setup. Nonetheless, the peak temperatures attained in such metathesis reactions are estimated to correlate with the boiling points of the byproduct alkali halide, sulphate, or nitrate salts.\cite {parkin1996solid} Techniques such as remote thermometry offers a promising method for accurately determining these temperatures in metathesis reactions—a subject beyond the scope of the current study.}


\section{\label{Section: Conclusions} Conclusions}

We have demonstrated the expansive applicability of solid-state metathesis reactions, not only in facilely accessing known $\rm Mg$- and $\rm Ca$-based materials under low temperatures \red{and ambient or vacuum pressures} but also in the discovering new materials with potential applications in energy storage. Specifically, through the metathetical approach, we serendipitously synthesised ilmenite-type $\rm Mg_3TeO_6$ (high-pressure, high-temperature polymorph) at low temperatures and ambient or vacuum pressure, eliminating the need for the megabar pressures conventionally required for its synthesis. Despite the sub-optimal pellet density, this ilmenite-type compound exhibits notable ionic conductivity at room temperature, designating ilmenite-structured materials as promising $\rm Mg^{2+}$ conductors. Furthermore, double-perovskite \blue {and garnet} compounds incorporating calcium, such as $\rm Ca_2ZnTeO_6$, display impressive ionic conductivity at room temperature, positioning them as competitive candidates for fast $\rm Ca^{2+}$ conduction. Leveraging the metathesis route, we also synthesised a diverse array of $\rm Mg$- and $\rm Ca$-based materials (including silicates, antimonates, ruthenates, and so forth), with potential \red{nanotechnology} applications beyond energy storage. A noteworthy aspect of the solid-state metathesis reactions is their adaptability to leverage lithium-, sodium-, and potassium-based functional materials \blue {(particularly alkali-rich layered, polyoxyanionic, spinel and rocksalt compounds)} for the low-temperature syntheses of calcium- and magnesium-based compounds \blue {(\textbf{Supplementary Figures 1, 7--12, 49, 50, 52, 59, 60}, and \textbf{Supplementary Tables 2--7})}, employing cost-effective salt precursors like \magenta {$\rm Ca(NO_3)_2$·$\rm 4H_2O$ (mineral analogue: nitrocalcite), $\rm CaCl_2$·$\rm 2H_2O$ (sinjarite), $\rm MgCl_2$ (chloromagnesite), $\rm MgSO_4$, $\rm MgCl_2$·$\rm 6H_2O$ (bischofite), and $\rm Mg(NO_3)_2$·$\rm 6H_2O$ (nitromagnesite)}. Future directions for improvement encompass shortening reaction times, presumably through the utilisation of microwave electromagnetic radiation, and optimising the quantity of salt precursors to tailor the nanomorphology of attainable $\rm Ca$- and $\rm Mg$-based materials. 
\red{Finally}, the versatility of solid-state metathesis reactions in obtaining new compound classes, characterised by intricate nanocrystal chemistry, offers valuable insights for the development of next-generation functional materials.

\newpage

\magenta {\section{\label{Section: Methods} Methods}

\subsection{Precursor compounds for metathesis reactions} 
Precursor compounds, which entailed lithium-, sodium-, potassium- and silver-based compounds, were prepared either by the conventional solid-state reaction synthesis route or the topochemical ion-exchange route, as detailed elsewhere\cite {kanyolo2023advances, he2017synthesis} \textbf{Supplementary Table 10} summarises the synthetic conditions used to attain various precursors.

\subsection{Metathesis reactions involving Ca salts} 
Unless otherwise stated, all protocols detailed herein were performed in a super dry room with a dew point maintained below --75$^\circ$C dP. Metathesis reactions to synthesise calcium-based compounds were conducted using nitrocalcite ($\rm Ca(NO_3)_2$·$\rm 4H_2O$; Sigma Aldrich, 99.0\%) or $\rm CaCl_2$·$\rm 2H_2O$ (Kishida Chemicals, 99\%) at 500 $^\circ$C under a nitrogen atmosphere. The heating temperature was determined based on (i) thermogravimetric measurements of the calcium salts (\textbf{Supplementary Figure 61}), indicating the temperature regime at which they are stable, and (ii) X-ray diffraction measurements of the reactions at various temperatures, 
indicating the temperature and duration (dwell time) necessary to trigger a complete metathesis reaction. 

The precursors ($\rm Li$-, $\rm Na$-, $\rm K$-, or $\rm Ag$-based compounds) were intimately mixed with $\rm Ca(NO_3)_2$·$\rm 4H_2O$ or $\rm CaCl_2$·$\rm 2H_2O$ using an agate mortar and pestle, and then placed in a gold or platinum crucible. Alumina crucibles could also be used, {\it albeit} posing some challenge in removing some fired powders that adhered to the bottom of the alumina crucibles. The mixture was heated at 500 $^\circ$C in a fixed nitrogen flux for 99 hours. The heating furnace was equipped with a fractional distillation system (Full-Tech, $\rm FT$-$\rm N2$-$\rm SP$) that automatically supplied a constant nitrogen flux maintained at 1 $\rm L$ $\rm min^{-1}$.

For instance, $\rm Ca_2MgTeO_6$ can be prepared via the following metathesis reaction using $\rm Ca$ salt at 500 $^\circ$C in a fixed nitrogen flux for 99 hours (\textbf{Supplementary Figure 62}):

\begin{align}
     \,{\rm Na_4MgTeO_6} + 2\,{\rm Ca(NO_3)_2} \cdot {\rm 4H_2O}  \ce{->} {\rm Ca_2MgTeO_6} + 4\,{\rm NaNO_3} + 8\,{\rm H_2O}\uparrow
\end{align}

\begin{align}
     \,{\rm Na_4MgTeO_6} + 2\,{\rm CaCl_2} \cdot {\rm 2H_2O}  \ce{->} {\rm Ca_2MgTeO_6} + 4\,{\rm NaCl} + 4\,{\rm H_2O}\uparrow
\end{align}

A slight molar excess (10\% mol) of $\rm Ca(NO_3)_2$·$\rm 4H_2O$ was utilised to ensure complete metathesis reaction. As for solid-state metathesis reactions involving $\rm CaCl_2$·$\rm 2H_2O$, a two-fold molar excess of $\rm CaCl_2$·$\rm 2H_2O$ was used. The resulting products, after completion of the metathesis reactions, were washed with ethanol (Kishida Chemicals, 99.5\%) under magnetic stirring to dissolve the remaining $\rm Ca(NO_3)_2$·$\rm 4H_2O$ and byproduct salts ({\it e.g.}, $\rm LiNO_3$, $\rm NaNO_3$, \textit {etc}.). Methanol could also be used, although we restricted ourselves to using ethanol due to the hazardous effects of handling methanol. It should also be noted that metathesis reactions involving $\rm K$-based compound precursors with $\rm Ca(NO_3)_2$·$\rm 4H_2O$ yield $\rm KNO_3$ as a byproduct, which is slightly soluble in ethanol. In this case, glycerol or distilled water could be used to completely dissolve the nitrate byproduct salt. Note that distilled water can only be used for metathesis reactions in which the target compounds are not hygroscopic. Thus, the choice of solvent depends on (i) the solubility of the byproduct ({\it i.e.}, alkali nitrates, chlorides, {\it etc}. (\textbf{Supplementary Table 11})), (ii) the solubility of $\rm Ca(NO_3)_2$·$\rm 4H_2O$ or $\rm CaCl_2$·$\rm 2H_2O$ (\textbf{Supplementary Table 12}) and (iii) the hygroscopicity of the target compounds. The target compound was finally washed with acetone, prior to drying for 10 hours in a vacuum oven at 100 $^\circ$C.

When the metathesis reaction (exemplified in equation (5)) was conducted at 500 $^\circ$C in air, instead of nitrogen, $\rm CaCO_3$ (vaterite) was observed as a minor impurity phase (\textbf{Supplementary Figure 63}). In addition to using $\rm Ca(NO_3)_2$·$\rm 4H_2O$, limited experiments were also performed with $\rm CaCl_2$·$\rm 2H_2O$ (Kishida Chemicals, 99\%), $\rm CaCl_2$ (Fujifilm Wako Pure Chemicals), and $\rm CaBr_2$ (Tokyo Chemical Industry (TCI), 97\%) (\textbf{Supplementary Figure 64}), indicating that hydrated calcium salts are effective for synthesising calcium-based compounds via such metathesis reactions.

Metathesis reactions of $\rm Ag$-based compounds with $\rm Ca(NO_3)_2$·$\rm 4H_2O$ or $\rm CaCl_2$·$\rm 2H_2O$ at 500 $^\circ$C yielded the target compound along with $\rm Ag$ metal as a byproduct (\textbf{Supplementary Figure 65}). This occurs because the $\rm AgNO_3$ byproduct, in principle, decomposes to $\rm Ag$ according to the following reaction:

\begin{align}
     2\,{\rm AgNO_3} \ce{->} 2\,{\rm Ag} + 2\,{\rm NO_2}\uparrow + \,{\rm O_2}\uparrow
\end{align}

Each typical batch yielded approximately 0.5 g of the target product. To increase the sample quantity, up to 10 batches were occasionally fired together in a single furnace session, resulting in a total yield of 5 g. Notably, depending on furnace capacity, metathesis reactions can be further optimised to produce larger quantities of the target powder on a pilot scale. A summary of the metathesis synthesis protocol for Ca-based compounds is illustrated in \textbf{Supplementary Figure 66}.

\subsection{Metathesis reactions involving Mg salts} 
Metathesis reactions to attain Mg-based compounds were also performed in a super dry room with a dew point maintained below --75$^\circ$C dP. Nitromagnesite ($\rm Mg(NO_3)_2$·$\rm 6H_2O$ (Kishida Chemicals)) and chloromagnesite ($\rm MgCl_2$ (Sigma-Aldrich, >98\%)) were mainly used in the metathesis reactions involving $\rm Li$-, $\rm Na$-, $\rm K$-, or $\rm Ag$-based compound precursors. 

For instance, $\rm Mg_3TeO_6$ can be prepared via the following metathesis reaction at 250$^\circ$C in air or nitrogen atmosphere for 99 hours:

\begin{align}
     \,{\rm Na_2Mg_2TeO_6} + \,{\rm Mg(NO_3)_2} \cdot {\rm 6H_2O}  \ce{->} {\rm Mg_3TeO_6} + 2\,{\rm NaNO_3} + 6\,{\rm H_2O}\uparrow
\end{align}

A twofold molar excess of $\rm Mg(NO_3)_2$·$\rm 6H_2O$ was utilised to ensure complete reaction. The precursors ($\rm Li$-, $\rm Na$-, $\rm K$-, or $\rm Ag$-based compounds) were intimately mixed with $\rm Mg(NO_3)_2$·$\rm 6H_2O$ using an agate mortar and pestle, and then placed in a gold, platinum or alumina crucibles. The resulting products, after completion of the metathesis reactions (at 250$^\circ$C in $\rm N_2$ for 99 hours), were washed with distilled water under magnetic stirring to dissolve the remaining $\rm Mg(NO_3)_2$·$\rm 6H_2O$ and byproduct nitrate salts ({\it e.g.}, $\rm NaNO_3$). The choice of solvent depends on (i) the solubility of the byproduct ({\it i.e.}, alkali nitrates, chlorides, {\it etc}. (\textbf{Supplementary Table 11})) and (ii) the solubility of Mg salt initially used (\textbf{Supplementary Table 13}). The target compound was finally washed with acetone, prior to drying for 10 hours in a vacuum oven at 100$^\circ$C. Subsequently, the target compound was reannealed at 400$^\circ$C for 24 hours under a nitrogen atmosphere (depending on the initial crystallinity) to enhance the material's crystallinity.

Metathesis reactions involving $\rm MgCl_2$ were performed at 400$^\circ$C for 99 hours under nitrogen or air. Just as is the case for metathesis reactions involving Ca salts, the heating temperature for metathesis reactions entailing Mg salts ($\rm MgBr_2$ (Fujifilm Wako Pure Chemicals), $\rm MgCl_2$ (Sigma-Aldrich, >98\%), $\rm MgBr_2$·$\rm 6H_2O$ (Fujifilm Wako Pure Chemicals, 99.9\%), $\rm MgSO_4$·$\rm 7H_2O$ (Kishida Chemicals, 99.5\%), $\rm MgCl_2$·$\rm 6H_2O$ (Kishida Chemicals, 98\%), $\rm MgSO_4$ (Kishida Chemicals)) was determined based on (i) thermogravimetric measurements of the magnesium salts (\textbf {Supplementary Figure 67}), indicating the temperature regime at which they are stable, and (ii) X-ray diffraction measurements of the reactions at various temperatures (\textbf {Supplementary Figures 68 and 69}), indicating the temperature and duration (dwell time) necessary to trigger the complete metathesis reactions.

Metathesis reactions to yield Mg-based compounds were found to be contingent on the firing atmosphere, precursors and Mg salts used, with some reactions (particularly those involving $\rm MgCl_2$) yielding trace to significant amounts of $\rm MgO$ (\textbf{Supplementary Figure 59a}). A summary of the metathesis protocol is illustrated in \textbf{Supplementary Figure 70}.

Additionally, for the sake of comparison, powder samples of ${\rm Mg}M_2{\rm TeO_6}$ ($M = \rm Co, Cu, Ni, Zn, Mg $), ${\rm Mg_2}M{\rm TeO_6}$, and ${\rm Mg_{1.5}}M_{1.5}{\rm TeO_6}$ were also prepared using high-temperature solid-state ceramics route (\textbf{Supplementary Figure 41}). \textbf{Supplementary Tables 1 and 2} show the annealing conditions utilised to attain Mg-based samples.
}

\subsection{X-ray diffraction (XRD) analyses}
Conventional XRD analyses were undertaken utilising a $\rm Bruker$ $\rm D8$ $\rm ADVANCE$ diffractometer to elucidate the crystallinity of \magenta {both the precursors and the powder samples attained via solid-state metathesis reactions.} The measurements were executed in Bragg-Brentano geometry mode employing Cu-{\it K}$\alpha$ radiation. Rietveld analyses were conducted for compounds from which meaningful structural data could be extracted (\textbf{Supplementary Figures 50 and 52}).

\subsection{Morphological and physicochemical characterisation}
\magenta {The chemical composition of the compounds attained via metathesis reactions were analysed by a scanning electron microscope ($\rm JSM$-$\rm 6510LA$) carried out using the energy dispersive X-ray ($\rm EDX$) imaging function. 
Stoichiometry quantifications, particularly of Ca-based compounds, were precisely determined by inductively coupled plasma mass spectrometry ($\rm ICP$-$\rm MS$).}
High-resolution scanning $\rm TEM$ ($\rm STEM$) imaging was executed utilising a $\rm JEOL$ $\rm JEM$-$\rm ARM200F$, equipped with a $\rm CEOS$ $\rm CESCOR$ $\rm STEM$ $\rm Cs$ corrector (spherical aberration corrector), under an acceleration voltage set at 200 kV. $\rm STEM$ images for $\rm CaMg_2TeO_6$, $\rm Ca_2MgTeO_6$ (\textbf{Figure {\ref {Figure 6}}}) and $\rm Ca_{1.5}Ni_2SbO_6$ (\textbf{Figure {\ref {Figure 8}}}) were acquired using a $\rm JEM$-$\rm ARM200F$ ($\rm NEOARM$), equipped with a $\rm CEOS$ $\rm ASCOR$ $\rm Cs$ corrector. Details pertaining to the sample preparation, are as detailed in ref.\citenum{masese2022honeycomb}. Electron microscopy measurements were systematically performed along various zone axes, specifically [100], [110], and [310] zone axes.

\subsection{Ionic conductivity measurements}

Powder samples of ${\rm Mg}M_2{\rm TeO_6}$, ${\rm Mg_2}M{\rm TeO_6}$, ${\rm Ca}M_2{\rm TeO_6}$, and ${\rm Ca_2}M{\rm TeO_6}$ ($M = \rm Mg, Zn, Ca $) underwent uniaxial pressing to form pellets with 10 mm diameters, applying a pressure of approximately 40 MPa. Stainless steel (SUS) served as the current collector for these pellets, with densities reaching around 50$\sim$60\% of their respective theoretical ceramic densities. Electrochemical impedance measurements (\textbf{Figures {\ref {Figure_6}a,b,e,f,g}} and \textbf {Supplementary Figure 57}) were conducted based on the protocols detailed elsewhere\cite {masese2022honeycomb}, \magenta {with careful measures taken to avoid exposing samples to moisture in the air.}

The activation energy ($E_{\rm a}$) for magnesium-ion or calcium-ion conduction was determined through linear fitting of bulk ionic conductivity values at different temperatures, employing the well-established Arrhenius equation as detailed in ref. \citenum {masese2022honeycomb}. EC-Lab software package Z-fit was used to fit all equivalent circuits of the Nyquist plots.

\subsection{Magnetic susceptibility measurements}
To assess the valency state of the transition metal atoms in the as-prepared ${\rm Mg}M_2{\rm TeO_6}$ \blue {($M = \rm Co, Ni $)} 
samples, magnetic susceptibility measurements (\textbf {Supplementary Figure 58}) were performed using a Quantum Design Magnetic Property Measurement System, at a temperature range of 2--300 K and a magnetic field range of $\pm 70$ kOe.

\subsection{Attenuated total reflectance Fourier-transform infrared spectroscopy ($\rm ATR$-$\rm FTIR$) measurements}
\magenta {$\rm ATR$-$\rm FTIR$ spectra were obtained using an $\rm IRTracer$-100 $\rm FTIR$ spectrometer (Shimadzu, Japan) equipped with a single-reflection $\rm ATR$ accessory featuring a diamond crystal ($\rm QATR$ 10). For each sample, 512 scans were averaged over the range of 4000 to 400 $\rm cm^{-1}$ with a resolution of 4 $\rm cm^{-1}$. Background scans were conducted prior to spectrum acquisition. The presence or absence of $\rm H_2O$ in the target samples was determined by analysing the characteristic absorption peaks observed in the $\rm ATR$-$\rm FTIR$ spectra.
Calcium- and magnesium-based compounds synthesised via solid-state metathesis reactions were subjected to $\rm ATR$-$\rm FTIR$ analysis to detect any traces of $\rm H_2O$ (\textbf {Supplementary Figure 56}). These analyses were benchmarked against reference compounds such as $\rm Mg(NO_3)_2$·$\rm 6H_2O$, $\rm H_6TeO_6$ (or equivalently as $\rm Te(OH)_6$), and $\rm H_2Mg_2TeO_6$. The $\rm H_2Mg_2TeO_6$ reference compound was synthesised through protonation of $\rm Na_2Mg_2TeO_6$ using 2 M acetic acid ($\rm CH_3COOH$), based on the following reaction: $\rm Na_2Mg_2TeO_6$ + $\rm 2CH_3COOH \ce{->} \rm 2CH_3COONa$ + $\rm H_2Mg_2TeO_6$. The absence of spectral features corresponding to $\rm H_2O$ (--$\rm OH$ functional groups) in the compounds obtained via solid-state metathesis reactions suggests that neither $\rm H^+$ nor $\rm OH^-$ ions were present, ensuring that the ionic conductivity measurements were not influenced by these species ({\it i.e.}, $\rm H^+$, $\rm OH^-$).}

\subsection{Nuclear magnetic resonance (NMR) measurements}
\magenta {$\rm ^{1}H$ magic-angle spinning ($\rm MAS$) $\rm NMR$ measurements (\textbf {Supplementary Figure 55}) were conducted at 298 K using a Bruker AV-300 spectrometer, operating at a $\rm ^{1}H$ Larmor frequency of 300 MHz, with a 1.4 mm $\rm MAS$ probe at a 40 kHz spinning speed. Spectra acquisition employed a rotor-synchronised Hahn echo sequence (90$^\circ$ - $\tau$ - 180$^\circ$ - $\tau$ - acquisition), and signal intensities were normalised based on the sample weights within the rotors. A recycle delay of 10 s was used to collect a total of 256 scans for each sample. $\rm ^{1}H$ chemical shifts were referenced to the $\rm CH_{3}COOLi$ resonance at --1.27 ppm.}

\subsection{Thermogravimetric measurements}
\magenta {Thermogravimetric and differential thermal analysis ($\rm TG$-$\rm DTA$) were performed using a $\rm Bruker$ $\rm AXS$ $\rm 2020SA$ $\rm TG$-$\rm DTA$ instrument in either nitrogen or air atmosphere. The samples were briefly exposed to air whilst being loaded into platinum crucibles prior to measurement. The analysis (\textbf {Supplementary Figures 61 and 67}) was conducted over a temperature range between 25 and 1000 $^\circ$C, at a heating rate of 5 $^\circ$C $\rm min^{-1}$.}

\subsection{Computational details}
\magenta {Density functional theory ($\rm DFT$) calculations were employed to evaluate the structural stability of the $\rm Mg_3TeO_6$ polymorphs, following the computational formalism outlined in previous studies.\cite {tada2022implications} Additional information regarding the methodology and parameters used is available in \textbf {Supplementary Note 1}.}

\newpage
\section*{Supplementary Information}
The online link to the Supplementary Information, which includes additional experimental details, will be made available in future updates of the preprint.

\newpage
\section*{Acknowledgements}
This work was supported by Green Technologies for Excellence (GteX) Program of the Japan Science and Technology Agency (JST). T.M. and G.M.K. would like to acknowledge the financial support from AIST Edge Runners Funding. \magenta {G.M.K. gratefully acknowledges the Japan Society for the Promotion of Science (JSPS) for the support of his fellowship.} S.K. acknowledges funding from the JST FOREST Grant (No. JPMJFR212V). \magenta {The computations in this work were conducted using the facilities of the Supercomputer Center at the Institute for Solid State Physics, University of Tokyo.}
\newpage

\section*{Conflicts of interest}
There are no conflicts to declare.

\newpage

\bibliography{rsc} 
\bibliographystyle{rsc} 

\end{document}